\documentclass[twocolumn,showpacs,preprintnumbers,amsmath,amssymb]{revtex4}
\usepackage{graphicx}% Include figure files
\usepackage{dcolumn}% Align table columns on decimal point
\usepackage{bm}% bold math

\begin{document}

\title{Theory of disordered flux-line liquids}

\author{A.M. Ettouhami} 
% \homepage{}
\email{mouneim@phys.ufl.edu}
\affiliation{Department of Physics, University of Florida, P.O. Box 118440,
Gainesville, FL 32611}

\date{\today}

\begin{abstract}

We study the equilibrium statics and nonequilibrium driven dynamics of 
flux line liquids in presence of a random pinning potential. 
Under the assumption of replica symmetry, we find 
in the static case using a replica Gaussian variational method 
that the only effect of disorder is to
increase the tilt modulus and the confining ``mass" of the internal modes of the
flux lines, thus decreasing their thermal wandering. 
In the nonequilibrium, driven case, we derive the long scale,
coarse-grained equation of motion of the vortices in presence of disorder, which apart from 
new Kardar-Parisi-Zhang nonlinearities, has the same form as the equation of motion
for unpinned vortices, with renormalized coefficients. 
This implies, in particular, that the structure factor of a disordered vortex liquid has the same 
functional form as in the absence of pinning, in disagreement with the results of previous
hydrodynamic methods. The expression of the static structure factor derived within our approach 
is consistent both with experimental data and with the standard theory of 
elasticity of vortex lattices.

\end{abstract}

\pacs{74.20.De, 74.25.Qt}

%\keywords{Vortices, vortex liquids, pinning of flux lines}

\maketitle

\section{Introduction}
\label{Introduction}

During the past fifteen years, the study of the properties of flux line liquids in
high temperature superconductors (HTSCs) has been one of the most active areas of research in vortex 
phenomenology. Yet, despite an impressive body of literature, both experimental and theoretical, and 
despite a relatively good understanding of the overall behavior and macroscopic properties of vortex 
liquids, it seems that the important question of the actual microscopic correlations of flux line 
trajectories inside such liquids has not been fully understood yet. Indeed, of the several 
theoretical approaches that have been used to study the properties of liquid vortex matter, 
one particular approach, which has had a rather strong impact on our present
understanding of flux liquids in HTSCs, is the boson mapping, developed by several authors,
  \cite{Nelson,Nelson-Seung,Nelson-LeDoussal,Tauber-Nelson,Tesanovic,Benetatos-Marchetti}
which is based on the observation   \cite{Fisher} that there is a formal mapping between the partition 
function of a three-dimensional system of interacting flux lines, and the 
imaginary-time partition function of quantum bosons in two-dimensions. 
Although the boson mapping is ultimately used to find density-density correlation functions, 
and does not contain, after coarse-graining, any detailed information about flux-line 
trajectories, it has been argued,   \cite{Nelson-Seung} based on the behavior of the structure 
factor derived in this and other hydrodynamic approaches, \cite{Radzihovsky-Frey} that flux lines 
wander throughout the sample in a random-walk-like fashion, much in the same way as in a 
hypothetical ``ideal gas" of noninteracting vortices. This implies, in particular, that the 
internal fluctuations of flux lines have an average spatial extent which diverges with the sample 
thickness $L$.

The above interpretation, and in fact the whole hydrodynamic approach to flux
line liquids, suffers from a number of inconsistencies which have been
pointed out and discussed in detail in two recent papers by the
author. \cite{Ettouhami1,Ettouhami2} In these two articles, the author has proposed a new approach to study
three-dimensional flux-line liquids in type II superconductors which,
instead of the density, uses the actual conformation variables of vortices as the
fundamental dynamical variables of the flux-line system.
This new approach, which makes contact with the standard theory of classical
fluids, is based on the separation of dynamical variables of flux lines into
center of mass (c.m. for short) and internal modes, and on the
observation that the repulsive interactions between flux lines must lead to a
certain degree of confinement of the internal modes, whose fluctuations are shown
to be bounded and no-longer diverge with the sample
thickness. \cite{Ettouhami1,Ettouhami2} This picture is obviously in contradiction
with the results of the boson mapping of 
refs.  \cite{Nelson-Seung},  \cite{Nelson-LeDoussal} and  \cite{Tauber-Nelson}.

In this paper we wish to generalize the methodology developed in these previous studies,
(refs.  \cite{Ettouhami1} and  \cite{Ettouhami2}), to study the statics and dynamics of 
vortex liquids in presence of a random pinning potential. In the static case, we shall show in 
particular that disorder leads to an enhancement of the tilt modulus and the confining 
``mass" of the internal modes of flux lines, thereby reducing their thermal wandering, but  
otherwise leaves the analytic form of the structure factor $S({\bf r},z)$ unchanged, 
contrarily to what has been argued by previous authors based on boson mapping 
methods \cite{Nelson-LeDoussal,Tauber-Nelson} or other hydrodynamic approximations.
 \cite{Radzihovsky-Frey}
These static results are then generalized to the dynamics of 
driven vortex liquids in the presence of a random pinning potential.
In contrast to earlier studies of this system, in which the density
was used as the fundamental dynamical variable, in our approach flux line
trajectories, which are the true dynamical variables of the system, are
used throughout. This enables us to derive the coarse-grained, large scale
equation of motion of vortices in the flux liquid in the presence of disorder, in
close analogy with earlier work on driven, disordered flux lattices. \cite{BMR,SV,LDG}

This article is organized as follows. In Sec. \ref{Disorder}, we use a simple Larkin
analysis to study the effect of disorder on vortex liquids. This Larkin analysis is then refined in
Sec. \ref{GVM} where we use the replica Gaussian variational method \cite{Mezard-Parisi} to properly
include the effect of the relatively strong thermal fluctuations which are a common characteristic of
HTSCs. In Sec. \ref{SecMSR}, we construct an action formulation for the dynamics of 
flux lines in a vortex liquid. Then, before considering the nonequilibrium case 
of a driven disordered flux liquid, we shall first be interested in the 
equilibrium dynamics of flux liquids, which we will investigate in quite 
some detail in Sec. V. Such an investigation is not only a natural step 
toward the more complex disordered case, but is also necessary for the
developments to follow, since one needs to correctly specify the 
near-equilibrium dynamics of the unpinned 
interacting liquid in order to be able to tackle the 
out-of-equilibrium driven, disordered case. In Sec. \ref{Hartree}, we 
derive the coarse-grained dynamics of driven, disordered flux liquids in 
the limit of high drives before deriving the structure factor of pinned flux liquids
in Sec. \ref{DynStrucFac}. Sec. \ref{Conclusion} contains 
a discussion of our results along with our conclusions.

\section{Flux line liquid in the presence of disorder: perturbative analysis}
\label{Disorder}

We thus consider a flux-line liquid in $d=d_\perp+1$ dimensions 
(we use $d_\perp=2$ in all explicit calculations), in presence of an
external pinning potential. Our starting point is the Hamiltonian: \cite{Nelson,Nelson-Seung}
\begin{eqnarray}
H & = & \sum_{i=1}^N\int \!\!dz\Big\{
\frac{1}{2}K\Big(\frac{d{\bf r}_i}{dz}\Big)^2 \!+\!
\frac{1}{2}\sum_{j(\neq i)} V\big({\bf r}_i(z)-{\bf r}_j(z)\big)\Big\}
\nonumber\\
& + & \sum_{i=1}^N\int dz\; V_d({\bf r}_i(z),z) \; ,
\label{def-H}
\end{eqnarray}
where the $d_\perp$-dimensional vector ${\bf r}_i(z)$ parametrizes the trajectory of the $i$-th flux-line
as it traverses the superconducting sample, $K$ is the tilt modulus of the flux lines, 
$V(r)=2\varepsilon_0K_0(r/\lambda)$ is the interaction potential between flux line elements at equal height,
and $V_d({\bf r},z)$ is a random pinning potential. In the above expression of $V(r)$, 
$\lambda$ is the London penetration depth in the $(ab)$ planes, 
$K_0$ is a modified Bessel function, \cite{Abramowitz}
and $\varepsilon_0=(\phi_0/4\pi\lambda)^2$, 
where $\phi_0=hc/2e$ is the flux quantum. \cite{deGennes} 
In equation (\ref{def-H}) and all equations below, the origin of heights is taken
to be at the center of the sample, and all $z$ integrals are taken from $-L/2$ to
$L/2$ ($L$ is the sample thickness). For simplicity, we shall consider
that the probability distribution of $V_d$ is Gaussian, with zero mean and variance
\begin{eqnarray} 
\langle V_d({\bf r},z)V_d({\bf r}',z')\rangle = \Delta({\bf r}-{\bf r}',z-z')\;. 
\end{eqnarray}
We next consider the canonical partition function of this system (here $T$ is temperature, and we use
units such that Boltzmann's constant $k_B=1$),
\begin{eqnarray}
Z = \int \prod_{i=1}^N[d{\bf r}_i(z)]\;\mbox{e}^{-H/T} \;,
\end{eqnarray}
and average over the disorder by introducing $p$ replicas of the above system and
making use of the well-known replica trick:
\begin{eqnarray} 
\overline{\ln Z} = \lim_{p\to 0}\frac{\overline{Z^p} - 1}{p} \; ,
\label{replica_trick}
\end{eqnarray}
upon which we obtain the following, disorder-averaged Hamiltonian:
\begin{eqnarray} 
\bar{H}  & \!=\! & \sum_{a=1}^p\sum_{i=1}^N\int\!\! dz\,\frac{1}{2}\big\{
K\Big(\frac{d{\bf r}_i^a}{dz}\Big)^2 \!\! +
\sum_{j(\neq i)} V\big({\bf r}_i^a(z)\!-\!{\bf r}_j^a(z)\big)\big\}
\nonumber\\ 
&-& \frac{1}{2T}\sum_{a,b=1}^p\sum_{i,j=1}^N\int\!\!dz\,dz'
\,\Delta\big({\bf r}_i^a(z) \!-\! {\bf r}_j^b(z');z-z'\big),
\label{Hbar}
\end{eqnarray}
where the superscripts $(a, b, \ldots)$ label replicas. 
In what follows, it will prove useful to write the flux line position at height $z$, 
${\bf r}_i(z)$, as the sum
\begin{eqnarray} 
{\bf r}_i(z) = {\bf r}_{0i} + {\bf u}_i(z) \; ,
\end{eqnarray}
where ${\bf r}_{0i}=\frac{1}{L}{\int} dz\;{\bf r}_i(z)$ is the c.m. position of the
$i$th flux line, 
while ${\bf u}_i(z)$ is the displacement of the $i$th flux line at height $z$ with
respect to ${\bf r}_{0i}$, and has the 
following decomposition into Rouse modes, \cite{Doi-Edwards}
\begin{eqnarray}
{\bf u}_i(z) = \sum_{n\neq 0} {\bf u}_i(q_n)\mbox{e}^{iq_nz} \;,
\end{eqnarray}
with the Fourier coefficients:
\begin{eqnarray}
{\bf u}_i(q_n) = \frac{1}{L}{\int} dz\;{\bf u}_i(z)\mbox{e}^{-iq_nz} \; .
\end{eqnarray}

In keeping with the spirit of the calculation carried out in ref. \cite{Ettouhami1}, 
in this section we shall perform a simple perturbative analysis of the physics encoded in the Hamiltonian 
(\ref{Hbar}) and expand $\bar{H}$ to quadratic order in the displacement field. The pure (disorder free) 
part of $\bar{H}$ yields:
\begin{eqnarray} 
\bar{H}_{pure} = \bar{H}_{pure}^{(0)} + \bar{H}_{pure}^{(1)}  \; ,
\label{Hbarpure}
\end{eqnarray}
where
\begin{eqnarray} 
\bar{H}_{pure}^{(0)} =  \frac{1}{2}\sum_{a=1}^p\sum_{i\neq j} LV({\bf r}_{0i}^a -
{\bf r}_{0j}^a)
\end{eqnarray}
is the Hamiltonian of a system of perfectly straight flux lines interacting
through the potential $V({\bf r})$, and where
\begin{eqnarray} 
\bar{H}_{pure}^{(1)} & = &\sum_{a=1}^p\sum_{i=1}^N\int dz\,
\frac{1}{2}\Big[
K\,\Big(\frac{d{\bf u}_i^a}{dz}\Big)^2 \!+\!
\mu_{\alpha\beta}^{(i)}u_{i\alpha}^a(z)u_{i\beta}^a(z)
\Big] 
\nonumber\\
&+& \sum_{a=1}^p\sum_{i=1}^N\sum_{j(\neq i)}\int dz
\;\frac{1}{2}\mu_{\alpha\beta}^{(ij)}u_{i\alpha}^a(z)u_{j\beta}^a(z)
\label{Hpure1}
\end{eqnarray}
represents the internal modes contribution to the ``pure" part. In the above
equation, the coefficients $\mu_{\alpha\beta}^{(i)}$ and $\mu_{\alpha\beta}^{(ij)}$
are given by:
\begin{eqnarray} 
\mu_{\alpha\beta}^{(i)} & = & \sum_{j(\neq i)=1}^N
\partial_\alpha\partial_\beta V({\bf r}_{0i}^a - {\bf r}_{0j}^a) \; ,
\label{Eq:mui}
\\
\mu_{\alpha\beta}^{(ij)} & = & -\partial_\alpha\partial_\beta 
V({\bf r}_{0i}^a - {\bf r}_{0j}^a) \; .
\label{Eq:muij}
\end{eqnarray}
In a similar fashion, a Taylor expansion of the disorder part of $\bar{H}$ to
quadratic order in the displacements 
$\{{\bf u}_i^a\}$ gives a decomposition similar to
the one in Eq. (\ref{Hbarpure}), namely
\begin{eqnarray} 
\bar{H}_{dis} = \bar{H}_{dis}^{(0)} + \bar{H}_{dis}^{(1)} \; . 
\end{eqnarray}
Here, $\bar{H}_{dis}^{(0)}$ is given by:
\begin{eqnarray} 
\bar{H}_{dis}^{(0)} = -\frac{1}{2T} \sum_{a,b}\sum_{i,j} 
L\bar\Delta({\bf r}_{0i}^a - {\bf r}_{0j}^b) \; ,
\end{eqnarray}
and is the disorder part of the disorder-averaged Hamiltonian of a system of
perfectly straight flux lines in a Gaussian random potential with variance
$\bar\Delta({\bf r})=\int_{-\infty}^{\infty}dz\,\Delta({\bf r},z)$. On the other hand, $\bar{H}_{dis}^{(1)}$
is given by
\begin{widetext}
\begin{eqnarray} 
H_{dis}^{(1)} & = & -\frac{1}{2T} \sum_{a=1}^p\sum_{i=1}^N\int dz\int dz'
\;\frac{1}{2}\big[u_{i\alpha}^a(z)-u_{i\alpha}^a(z')\big]
\big[u_{i\beta}^a(z)-u_{i\beta}^a(z')\big]\partial_\alpha\partial_\beta\Delta({\bf 0},z-z') + 
\nonumber\\
& - & \frac{1}{2T} \sum_{a=1}^p\sum_{i\neq j}\int dz\int
dz'\;\frac{1}{2}\big[u_{i\alpha}^a(z)-u_{j\alpha}^a(z')\big]
\big[u_{i\beta}^a(z)-u_{j\beta}^a(z')\big]\partial_\alpha\partial_\beta\Delta({\bf r}_{0i}^a-{\bf r}_{0j}^a,z-z') + 
\nonumber\\   
& - & \frac{1}{2T} \sum_{a\neq b}\sum_{i,j}\int dz\int dz'\;\frac{1}{2}\big[u_{i\alpha}^a(z)-u_{j\alpha}^a(z')\big]
\big[u_{i\beta}^b(z)-u_{j\beta}^b(z')\big]
\partial_\alpha\partial_\beta\Delta({\bf r}_{0i}^a-{\bf r}_{0j}^b,z-z') \; .
\label{Hexpansion}
\end{eqnarray}
\end{widetext}\noindent 
It is not difficult to see that the first term on the right hand side of the
above equation represents same replica, single-line contributions to the 
Hamiltonian of the internal modes of vortices, while the second and third terms represent 
contributions to $\bar{H}_{dis}^{(1)}$ coming from same replica, different flux
lines and from different replicas, respectively.

Collecting all terms, it follows that $\bar{H}$ can be written in the form
\begin{eqnarray} 
\bar{H} = \bar{H}^{(0)} + \bar{H}^{(1)}  \; ,
\end{eqnarray}
where 
\begin{subequations}
\begin{eqnarray} 
\bar{H}^{(0)} = \bar{H}^{(0)}_{pure} + \bar{H}^{(0)}_{dis} \; ,
\\
\bar{H}^{(1)} = \bar{H}^{(1)}_{pure} + \bar{H}^{(1)}_{dis} \; .
\end{eqnarray}
\end{subequations}
In the spirit of ref.  \cite{Ettouhami1}, we shall derive an effective
Hamiltonian for the internal modes of the flux lines by averaging 
$\bar{H}^{(1)}$ over the center of mass positions $\{{\bf r}_{0i}^a\}$:
\begin{eqnarray} 
H_{u} = \langle \bar{H}^{(1)}\rangle_0 = 
\langle \bar{H}^{(1)}_{pure}\rangle_0 + \langle \bar{H}^{(1)}_{dis}\rangle_0 \;,
\end{eqnarray}
where the average is carried out with statistical  
weight $\exp(-\bar{H}^{(0)}/T)/Z_0$ (with
$Z_0=\mbox{Tr}(\mbox{e}^{-\bar{H}^{(0)}/T})$).
The pure part of $H_{u}$ has already been evaluated in ref.  \cite{Ettouhami1}, 
with the result:
\begin{eqnarray} 
H_{u}^{pure} & = & \sum_{a=1}^p\Big\{
\sum_{i=1}^N\int \!\!dz\;\frac{1}{2}\Big[
K\,\Big(\frac{d{\bf u}_i^a}{dz}\Big)^2 + 
\mu\,{\bf u}_{i}^a(z)\cdot{\bf u}_{i}^a(z)
\Big] 
\nonumber\\
& + & \sum_{i=1}^N\sum_{j(\neq i)}\int dz
\;\frac{1}{2}\,\frac{\mu}{N-1}\,{\bf u}_{i}^a(z)
\cdot{\bf u}_{j}^a(z)
\Big\} \; ,
\end{eqnarray}
where the ``mass'' coefficient $\mu$ is given by
\begin{eqnarray}
\mu = \frac{\rho}{d_\perp}\int d{\bf r}\; g_0({\bf r})\nabla_\perp^2V({\bf r}) \;,
\label{mu}
\end{eqnarray}
and where
\begin{eqnarray}
g_{0}({\bf r}-{\bf r}') = \frac{1}{\rho^2}\sum_{i=1}^N\sum_{j(\neq i)}\langle
\delta({\bf r}-{\bf r}_{0i}^a)\delta({\bf r}'-{\bf r}_{0j}^a)
\rangle_0  
\end{eqnarray}
is the pair distribution function of the two-dimensional liquid formed by the centers of mass of
flux lines belonging to the same replica in the vortex liquid.
In a similar fashion, we show in Appendix \ref{App:Heff} that $H_{u}^{dis}=\langle
\bar{H}^{(1)}_{dis}\rangle_0$ can be written in the form:
\begin{eqnarray}
&H_{u}^{dis}&  =  \sum_{a=1}^p\sum_{i=1}^N\int dz\,\Big[
\frac{1}{2}\delta{K}(\partial_z{\bf u}_i^a)^2 \!+\! 
\frac{1}{2}\delta\mu\big({\bf u}_i^a(z)\big)^2
\Big] +
\nonumber\\
& + & \sum_{a=1}^p\sum_{i\neq j}\int dz\int dz'\; \delta\mu_{ij}^{(a)}(z-z')\,{\bf
u}_i^{a}(z)\cdot{\bf u}_j^a(z') +
\nonumber\\
& + & \sum_{a\neq b}\sum_{i, j}\int dz\int dz'\; 
\delta\mu_{ij}^{(ab)}(z-z')\,{\bf u}_i^{a}(z)\cdot{\bf u}_j^b(z') \; .
\nonumber\\
\label{Heff_dis}
\end{eqnarray} 
Here the long-wavelength disorder contribution $\delta{K}$ to the tilt modulus of
the flux lines is given by
\begin{eqnarray}
\delta K = -\frac{1}{d_\perp T}\int_{-\infty}^\infty
dz\; z^2\nabla_\perp^2\Delta({\bf r},z)\Big|_{{\bf r}=0} \; ,
\label{deltaK1}
\end{eqnarray}
and the ``mass'' coefficients $\delta\mu$ are given by
\begin{subequations}
\begin{eqnarray}
\delta\mu & = & -\frac{\rho}{d_\perp T}\int d{\bf r}\;\sum_{b=1}^p 
g_{0,ab}({\bf r}) \nabla_\perp^2\bar\Delta({\bf r}) \,,
\nonumber\\
\label{del_mu}\\ 
\delta\mu_{ij}^{(a)}(z) & = &\frac{\rho}{(N-1)d_\perp T}\int d{\bf r}\;g_0({\bf r})
\nabla_\perp^2\Delta({\bf r},z) \,,
\nonumber\\
\label{mu2}\\
\delta\mu_{ij}^{(ab)}(z) & = & \frac{\rho}{Nd_\perp T}\int d{\bf r}\;g_{0,a\neq b}({\bf r})
\nabla_\perp^2\Delta({\bf r},z) \,.
\nonumber\\
\label{mu3}
\end{eqnarray} 
\end{subequations}
We immediately note that the two-body coefficients $\delta\mu_{ij}^{(a)}(z)$ and 
$\delta\mu_{ij}^{(ab)}(z)$ vanish in the thermodynamic ($N\to\infty$) limit.
In equation (\ref{del_mu}) above, the sum $\sum_{b=1}^p 
g_{0,ab}({\bf r})$ denotes the quantity:
\begin{eqnarray}
\sum_{b=1}^p g_{0,ab}({\bf r}) = g_0({\bf r}) +\sum_{b(\neq a)=1}^p
g_{0,a\neq b}({\bf r}) \; ,
\end{eqnarray}
where
\begin{eqnarray}
g_{0,a\neq b}({\bf r}-{\bf r}') = \frac{1}{\rho^2}\sum_{i=1}^N\sum_{j=1}^N\langle
\delta({\bf r}-{\bf r}_{0i}^a)\delta({\bf r}'-{\bf r}_{0j}^b)
\rangle_0
\label{g0_aneqb}
\end{eqnarray}
is the pair distribution function of the c.m. mode of flux lines from different
replicas. Now, in Eq. (\ref{del_mu}), and under the {\em assumption}
of replica symmetry, all pair distribution functions $g_{0,a\neq b}({\bf r})$ are
equal to the same function $\tilde{g}_0({\bf r})$, and we 
may replace the sum $\sum_{b\neq a}g_{0,a\neq b}$ by $(p-1)\tilde{g}_{0}$, with
Eq. (\ref{del_mu}) becoming:
\begin{eqnarray}
\delta\mu = -\frac{\rho}{d_\perp T}\int d{\bf r}\;\Big(
g_0({\bf r}) + (p-1)\tilde{g}_{0}({\bf r})\Big)\nabla_\perp^2\bar\Delta({\bf r}),
\nonumber
\end{eqnarray}
which, in the limit $p\to 0$ reduces to:
\begin{eqnarray}
\delta\mu\big|_{p\to 0} = -\frac{\rho}{d_\perp T}\int d{\bf r}\;\Big(
g_0({\bf r}) - \tilde{g}_{0}({\bf r})\Big)\nabla_\perp^2\bar\Delta({\bf r})\,.
\end{eqnarray}
In the replica-symmetric ground state considered here,
the diagonal and off-diagonal (in replica space) pair distribution functions 
should be equal,
\begin{eqnarray}
g_0({\bf r}) = \tilde{g}_{0}({\bf r}) \;,
\end{eqnarray}
and we therefore obtain that the correction $\delta\mu$ identically vanishes,
which shows that the bare mass $\mu$ generated by interactions between flux lines
is unrenormalized by disorder.

We now can write the following expression for the effective Hamiltonian 
$H_{u} = H_{u}^{pure} + H_{u}^{dis}$ of the internal fluctuations of flux lines in 
a vortex liquid:
\begin{eqnarray}
H_{u} \simeq \sum_{a=1}^p\sum_{i=1}^N\int \!\!dz \;\frac{1}{2}\Big[
K_R\,(\partial_z{\bf u}_i^a)^2 + \mu\,\big({\bf
u}_i^a(z)\big)^2
\Big]  ,
\label{result-Heff}
\end{eqnarray}
where we discarded the terms proportional to $1/N$ (which vanish in the 
thermodynamic limit, see Eqs. (\ref{mu2})-(\ref{mu3})), and where the
renormalized tilt modulus $K$ is given by
\begin{eqnarray}
K_R = K + \delta K \; ,
\end{eqnarray}
with $\delta K$ given by Eq. (\ref{deltaK1}).
At this stage, an explicit expression for the disorder correlator 
$\Delta({\bf r},z)$ is called for.
For point disorder such as oxygen vacancies in HTSCs, 
we shall take $\Delta({\bf r},z) = \Delta_0\;\exp(-(r^2+z^2)/2\xi^2)$,
(with the understanding that the correlation length $\xi$ is
much smaller that the average intervortex distance $a=1/\sqrt{\rho}$),
upon which we obtain the following expression for the effective tilt modulus:
\begin{eqnarray}
K_R = K + \frac{\sqrt{2\pi}\Delta_0\xi}{d_\perp T} \; .
\label{KRtaylor}
\end{eqnarray}
This expression shows that a flux line liquid in presence of a weak pinning potential is 
equivalent to an unpinned liquid but with a higher tilt modulus, i.e. that flux lines are 
stiffened by weak point disorder, which is what one would expect based on physical intuition. 

For disorder that is correlated along the direction of flux
lines, e.g. columnar pins, \cite{Konc} the variance $\Delta({\bf r},z)$ is
$z$-independent. 
Upon using for this case a disorder correlator of the form
$\Delta({\bf r},z) = \Delta_0\;\exp(-r^2/2\xi^2)$,
we obtain
\begin{eqnarray}
K_R = K + \frac{L^3\Delta_0}{3d_\perp T\xi^2} \; ,
\end{eqnarray}
which shows that the tilt modulus is much strongly renormalized by correlated
disorder than it is by ordinary point disorder, in agreement with the predictions
of refs. \cite{Nelson-Vinokur,GLD}. The divergence of the right hand side
of the above equation, however, signals the breakdown of perturbation theory for 
correlated disorder, which is best treated with other, 
nonperturbative methods \cite{Nelson-Vinokur} that are better suited to strong
pinning situations.

\section{Variational approach}
\label{GVM}

We now generalize the analysis of the previous Section to take into account the
effect of possible large fluctuations of flux line trajectories, by using the replica
Gaussian variational approach for elastic manifolds. \cite{Mezard-Parisi}
For compactness, we shall only give the salient features of the calculation,
and refer the reader interested in more details to ref. \cite{Ettouhami2},
where a similar calculation was done for the pure case, the generalization to the
disordered case being straightforward.

We start by introducing the following variational Hamiltonian:
\begin{eqnarray}
H_v = H_0[\{ {\bf r}_{0i}\}] + H_1[\{ {\bf u}_{0i}\}] \; ,
\label{decomp}
\end{eqnarray}
where $H_0$ and $H_1$ are trial Hamiltonians for the c.m. and internal modes, 
respectively, and are to be determined variationally.
Although one can, in principle, use a very general trial Hamiltonian for the
internal modes of flux lines of the form
\begin{equation}
H_1 = \sum_{a,b}\sum_{i,j}\int\!\! dz\,dz'\; 
[G^{-1}(q_n)]_{ij,ab}^{\alpha\beta} u_{ia}^{\alpha}(q_n) 
u_{jb}^{\beta}(-q_n) ,
\end{equation}
the insight we gained from the perturbative solution suggests the following,
simplified form:
\begin{eqnarray}
H_1 = \sum_{a,i}\sum_{n\neq 0}
G^{-1}(q_n) |{\bf u}_{ia}(q_n)|^2 \; .
\label{defH1}
\end{eqnarray}
Variation of the trial free energy 
\begin{equation}
F_1=-T\ln Z_1+\langle \bar{H} -H_v\rangle_1 \; ,
\end{equation}
where $Z_1=\mbox{Tr}(\exp(-H_1/T))$ and $\langle\cdots\rangle_1$ denotes averaging
with statistical weight $\exp(-H_1/T)/Z_1$, with respect to the
c.m. Hamiltonian ${H}_0$ leads to the result that the optimal
choice for $H_0$ is given by
(we henceforth use the
shorthand notation $\int_{\bf q}=\int\frac{d^{d_\perp}\bf q}{(2\pi)^{d_\perp}}$):
\begin{eqnarray}
\tilde{H}_0 & = &\frac{L}{2}\sum_{i,j}\Big\{\sum_{n\neq 0} d_{\perp} pT
\Big[G_0^{-1}(q_n) - G^{-1}(q_n)\Big]\,G(q_n)
\nonumber\\
&+&\sum_{a=1}^p\int_{\bf q} V({\bf q})\,\mbox{e}^{i{\bf q}\cdot
({\bf r}_{0i}^a-{\bf r}_{0j}^a)}\,\mbox{e}^{-\frac{q^2}{2d_\perp}\phi_{ij}(0)}
\nonumber\\
& - & \frac{1}{T}\;\sum_{a,b}\int_{\bf q}\int dz\; \Delta({\bf q},z)
\,\mbox{e}^{i{\bf q}\cdot({\bf r}_{0i}^a-{\bf r}_{0j}^b)}
\,\mbox{e}^{-\frac{q^2}{2d_\perp}\phi_{ij}(z)} 
\Big\},
\nonumber\\
\label{H10}
\end{eqnarray}
where we defined
$\phi_{ij}(z)=\langle[{\bf u}_i(z)-{\bf u}_j(0)]^2\rangle$.
Further variation of the resulting free energy 
$F_v= -T\ln Z_1|_{H_0=\tilde{H}_0}$
with respect to $G(q_n)$ leads to the following result:
\begin{widetext}
\begin{eqnarray}
\tilde{G}^{-1}(q_n) & = &  q_n^2 - \frac{\rho}{d_\perp}\int_{\bf q} q^2 
V({\bf q})g_0({\bf q}) \mbox{e}^{- q^2 G(0)}
+ \frac{1}{d_\perp T}\int_{\bf q}dz\,q^2\Delta({\bf q},z)[1-\cos(q_nz)]
\mbox{e}^{-q^2\phi(z)/2d_\perp}
\nonumber\\  
& + &  \frac{\rho}{2}\int_{\bf q}dz\,q^2\Delta({\bf q},z)\,\Big(g_0(q)
+ (p-1)\tilde{g}_0(q)\Big)
\mbox{e}^{-q^2\phi(z)/2d_\perp}\,,
\end{eqnarray}
\end{widetext}\noindent
where now $\phi(z)$ denotes the relative displacement of internal modes within the 
same flux-line, $\phi(z)=\langle[{\bf u}_i(z)-{\bf u}_i(0)]^2\rangle$, and
where the tilde indicates that $\tilde{G}^{-1}(q_n)$ has been averaged over
all possible configurations of the c.m. positions $\{{\bf r}_{0i}^a\}$ which are
compatible with a liquid structure. \cite{Ettouhami1,Ettouhami2}
Under the assumption of replica symmetry (which implies
that $\tilde{g}_0(q)=g_0(q)$), the last term vanishes again in the
limit $p\to 0$, and we obtain:
\begin{eqnarray}
\tilde{G}^{-1}(q_n) & = &  q_n^2 - \frac{\rho}{d_\perp}\int_{\bf q} q^2
V({\bf q})g_0({\bf q})
\mbox{e}^{- q^2 \langle u^2\rangle/d_\perp} +
\nonumber\\
& \!+\! & \frac{1}{d_\perp T}\int_{\bf q}\!\!dz\,q^2\Delta({\bf q},z)[1-\cos(q_nz)]
\mbox{e}^{-\frac{q^2}{2d_\perp}\phi(z)}.
\nonumber\\
\label{resGa}
\end{eqnarray}
The last term on the right hand side of the above expression leads to the
following renormalized value of the tilt 
modulus of flux lines in the long wavelength limit: 
\begin{eqnarray}
K_R = K +\frac{1}{d_\perp T}\int_{\bf q}\int dz\; q^2z^2\Delta({\bf q},z)
\mbox{e}^{-q^2\phi(z)/2d_\perp} \; .
\label{K_R}
\end{eqnarray}
We thus see that the inverse propagator for the elastic distortions of flux lines
in the vortex liquid is given by the following, generic form:
\begin{eqnarray}
\tilde{G}^{-1}(q_n) = L(K_R q_n^2 + \mu_R) \; ,
\label{genericform}
\end{eqnarray}
where the effective ``mass" coefficient of internal modes
fluctuations $\mu_R$ is identical to the quantity derived in
ref. \cite{Ettouhami2}~:
\begin{eqnarray}
\mu_R = -\frac{\rho}{d_\perp}\int_{\bf q} q^2V({\bf q})g_0({\bf q})  
\mbox{e}^{-q^2 \langle u^2\rangle/d_\perp} \,,
\label{defmu}
\end{eqnarray}
except that it now depends on $K_R$ (through $\langle u^2\rangle$) 
and hence on the disorder strength $\Delta_0$.

In Eq. (\ref{K_R}), the relative displacement of the internal modes of a given flux line
\begin{subequations}
\begin{eqnarray}
\phi(z) & = & d_\perp T\sum_{n\neq 0}\tilde{G}(q_n)[1-\cos(q_nz)] \,,
\label{defphi2}\\
& = & \frac{d_\perp T}{\sqrt{{\mu}K_R}}
\,\Big[ 1 - \exp\big(-\sqrt{\mu/K_R}|z|\big)\Big]\,
\label{phiofz}
\end{eqnarray}
\end{subequations}
depends on $K_R$, and we therefore see that Eq. (\ref{K_R}) is in fact a
self-consistent equation for the effective tilt modulus. 
For the explicit evaluation of $\mu_R(T)$ and $K_R$, we shall make use of
the analytical ansatz of ref.  \cite{Ettouhami1} for the pair correlation
function $g_0({\bf r})$, which is given by:
\begin{eqnarray}
g_0({\bf r}) = 1 - \eta\exp(-{\alpha}r^2/a^2) \; ,
\label{g0}
\end{eqnarray}  
where $\alpha$ is a constant of order unity, and $0<\eta<1$. The numerical constant
$\eta$ quantifies the degree of correlation between c.m. positions of flux lines.
It is close to unity when flux lines are strongly anti-correlated due to the
repulsive interactions between their surrounding supercurrents, and close to zero
in situations where there is considerable cutting and crossing of flux lines. Using
the above ansatz for $g_0(r)$, we obtain the following expression for the ``mass''
$\mu_R$ of the internal modes as a function of $T$, \cite{Ettouhami2}
\begin{equation}
\frac{\mu_R(T)}{\mu_0} =
\left( \sqrt{1+\Big(\frac{\alpha T}{4a^2\sqrt{K_R\mu_0}}\Big)^2}
- \Big(\frac{\alpha T}{4a^2\sqrt{K_R\mu_0}}\Big)\right)^2 \,,
\label{muvsT}
\end{equation}
where $\mu_0=\mu_R(T=0)=2\eta\pi\rho\varepsilon_0/d_\perp$. Since $\mu_R$ is a
monotonically decreasing function \cite{Ettouhami2} of the parameter 
$\nu=(\alpha T/4a^2\sqrt{K_R\mu_0})$, 
we arrive at the important conclusion that pinning disorder (which increases
the value of the tilt modulus from its bare value $K$ to the effective value
$K_R>K$) {\em increases} the value of $\mu$, thereby reducing even further (than
the sole increase in $K$) the thermal wandering of flux lines. This
effect, which did not appear in the elementary treatment of Sec. \ref{Disorder}, is
expected to be rather weak for the weak disorder considered in this work, but may
nevertheless reveal itself in actual experiments.

Going back to the effective tilt modulus of the pinned flux line liquid, we see that 
due to the highly nonlinear character of the self-consistency equation (\ref{K_R}), 
it is not possible to solve this
equation for $K_R$ and obtain a general expression for the effective tilt
modulus in closed analytic form. For weak disorder, however, such that
$\Delta_0\ll K T/\xi$, one can evaluate the second term in
Eq. (\ref{K_R}) perturbatively in $\Delta_0$, by using
for $\phi(z)$ its expression in the absence of disorder, Eq. ({\ref{phiofz}}), with
$K$ instead of $K_R$. In the limit of small correlation lenght $\xi$, we obtain
\begin{eqnarray}
K_R = K \,\Big(
1 + \frac{\sqrt{\pi}\Delta_0K\xi^3}{2\sqrt{2}d_{\perp} T^2}
\Big) \,.
\label{resKR}
\end{eqnarray}
Comparison with Eq. (\ref{KRtaylor}) shows that the correction to the bare tilt
modulus $K$ goes to zero at high temperature more rapidly than in the simple Taylor
result of Eq. (\ref{KRtaylor})
($1/T^2$ as opposed to $1/T$). This discrepancy is due to the fact that
the Taylor expansion of Sec. \ref{Disorder} does not take proper account of
thermal fluctuations of the internal modes of flux lines, as opposed to the
self-consistent approach of this Section which in fact can be 
shown \cite{Mezard-Parisi} to become exact in the limit $d_\perp\to\infty$.

\section{Action formulation of the dynamics of flux-line liquids}
\label{SecMSR}

We now turn our attention to the dynamics of vortex liquids.
The general dynamical behavior of flux line assemblies (solids and liquids) 
in a random pinning environment has
attracted a lot of attention in recent years due on the one hand to the
considerable technological implications of understanding the flow of vortices
in HTSCs, and on the other hand because of
the fundamental theoretical questions and variety of physical regimes
displayed by these systems. Most of the above mentioned attention has focused
on the dynamics of flux solids, with questions about the degree of
crystalline and temporal order in the driven regime, and glassiness in the
absence of external drive, at the forefront of theoretical issues that have
been addressed. Although there have been a number of studies of the dynamics
of disordered flux liquids, these studies were either done
within hydrodynamic approaches which, by definition (since they involve a
coarse graining procedure over many vortices) are unable to give information
about actual flux-line trajectories inside the superconducting sample,
or used qualitative arguments to separate length and time scales in
the plastic regime near the melting point. \cite{Blatter-et-al}
While the latter approach is very valuable
in that it helps draw a good qualitative picture of the physics
of driven flux liquids, it assumes that the vortex liquid is very viscous and hence
only applies very close to the melting point. As we mentioned in the Introduction, here
our goal is to go beyond these previous treatments, and establish a general framework for 
a systematic perturbative study of driven flux-line liquids
in presence of weak disorder.

We shall assume that the motion of flux lines in the liquid state 
in presence of an external driving force ${\bf F}$ is
governed by the following, overdamped Langevin equation:
\begin{equation}
\gamma \partial_t {\bf r}_i(z,t)  = -\frac{\delta H}{\delta {\bf r}_{i}(z,t)} +
{\bf F} + {\bf\zeta}_{i}(z,t) \,.
\label{langevin}
\end{equation}
For simplicity, the distribution of the thermal noise $\zeta_i(z,t)$ will be taken
to be Gaussian, with zero mean and correlations (we use units such that $k_B=1$):
\begin{equation}
\langle\zeta_{i\alpha}(z,t)\zeta_{j\beta}(z',t')\rangle = 2\gamma 
T\delta_{ij}\delta_{\alpha\beta}\delta(z-z')\delta(t-t') .
\end{equation}
In the above equations, the parameter $\gamma$ is the microscopic friction
coefficient characteristic of the interaction of the system with the degrees
of freedom of the surrounding heat bath. In our case of flux lines in a type
II superconductor, $\gamma$ describes the drag on a flux line due to the
interactions of the electrons in the normal vortex cores with the 
underlying solid, and is given by the Bardeen-Stephen
expression: \cite{Bardeen-Stephen,deGennes}
\begin{eqnarray}
\gamma = \frac{\rho h^2}{8\pi e^2\xi_{ab}^2}\;\sigma_n \,,
\end{eqnarray}
where $\rho$ is the average density of flux lines, $\xi_{ab}$ is the
superconducting coherence length in the (ab) planes, 
and $\sigma_n$ is the normal-state conductivity.

The dynamics represented by Eq. (\ref{langevin}) is best studied using the
action formulation of Martin, Siggia and 
Rose \cite{MSR,Janssen,DeDominicis,Bausch} (MSR), whereby disorder-averaged
observables are obtained from the following, disorder-averaged generating
functional: \cite{Radzihovsky-Frey}
\begin{eqnarray}
{\cal Z} = \int \prod_{i=1}^N[d{\bf r}_i(z,t)]
[d\tilde{\bf r}_i(z,t)]\;\mbox{e}^{-\cal A} \,.
\label{MSR}
\end{eqnarray}
Here, the MSR dynamical ``action'' can be written in the form:
\begin{eqnarray}
{\cal A} & = & {\cal A}_{free} + {\cal A}_{int} + {\cal A}_{dis} \,,
\label{decompA}
\end{eqnarray}
where the ``free'' part
\begin{eqnarray}
{\cal A}_{free} & = &\sum_{i=1}^N\int dz\, dt 
\;\Big\{\frac{1}{2}\,(2\gamma T)\tilde{\bf r}_i^2(z,t)
\nonumber\\
&+& i\tilde{\bf r}_i(z,t)\cdot\big[\,\gamma\partial_t{\bf r}_i(z,t) 
- K\partial_z^2{\bf r}_i(z,t)
\big]\Big\}
\label{Afree}
\end{eqnarray}
corresponds to an ``ideal gas'' of non-interacting flux lines, and
\begin{equation}
{\cal A}_{int}  =  \sum_{i\neq j}\int dz\, dt\;i\tilde{\bf r}_i(z,t)\cdot\nabla
V({\bf r}_i(z,t)-{\bf r}_j(z,t))
\label{Aint}
\end{equation}
is the part of the action describing the interactions between vortices. The last term 
in Eq. (\ref{decompA}) decribes the pinning of flux lines by the underlying disorder potential, and is 
given by:
\begin{eqnarray}
{\cal A}_{dis} & = & -\frac{1}{2}\sum_{n,m}\int\!dzdt\int\!dz'dt'
\;\tilde{\bf r}_{n\alpha}(z,t)\tilde{\bf r}_{m\beta}(z',t')
\nonumber\\
&\times&\partial_\alpha\partial_\beta
\Delta\big({\bf r}_n(z,t) - {\bf r}_m(z',t')\big)\,.
\end{eqnarray}
(In the above equations, and in what follows, $\sum_{i\neq j}$ 
stands for the double summation $\sum_{i=1}^N\sum_{j(\neq i)}$).
Our main goal in the next two Sections will be to find a way to calculate
expectation values of dynamical observables in the liquid phase (like for
example the dynamic structure factor of the flux line liquid) by 
integrating directly over the conformation variables 
$\{\tilde{\bf r}_{i}(z,t),{\bf r}_{i}(z,t)\}$ using the MSR generating functional of Eq. 
(\ref{MSR}), and not by integrating over the density operators, as is done 
in the (static) boson analogy  \cite{Nelson-Seung} and other hydrodynamic
approaches. \cite{Radzihovsky-Frey} This means that we have to
avoid writing the dynamical action ${\cal A}$ in terms of the density
operator 
$\hat\rho({\bf r},z;t)=\sum_{i=1}^N\delta\big({\bf r}- {\bf r}_i(z,t)\big)$,
and instead keep the conformation variables 
$\{\tilde{\bf r}_i(z,t),{\bf r}_i(z,t)\}$ as the true and only dynamical
variables in the problem. Our strategy will be very similar to the strategy
adopted in the previous two Sections, which 
consists in separating out the center of mass from the internal modes of
flux lines, and trying to find a decoupled approximation to the dynamic
action ${\cal A}$
\begin{equation}
{\cal A} = {\cal A}^{(0)} + {\cal A}^{(u)} \,,
\label{decomposition}
\end{equation}
such that ${\cal A}^{(0)}$ and ${\cal A}^{(u)}$ depend only on the c.m. and
on the internal modes, respectively. Since our ability to perform functional
integrations is limited to Gaussians, for a decomposition of 
the form (\ref{decomposition}) to be useful at all we will need to write 
${\cal A}^{(u)}$ as a bilinear form in the displacement fields 
$\{\tilde{\bf u}_{i}(z,t),{\bf u}_{i}(z,t)\}$ fields. 
Such a decomposition will allow us to evaluate averages of the form:
%\begin{widetext}
\begin{eqnarray}
\langle \hat\rho({\bf r},z;t)\hat\rho({\bf r'},z';t')\rangle  &=&   
\sum_{n,m}\big\langle\delta\big({\bf r}-{\bf r}_{0n}(t)-{\bf u}_n(z,t)\big)
\nonumber\\
&\times&\delta\big({\bf r}'-{\bf r}_{0m}(t')-{\bf u}_m(z',t')\big)\big\rangle
\nonumber\\
\label{avg-density}
\end{eqnarray}
%\end{widetext}\noindent
by integrating over the c.m. modes $\{{\bf r}_{0i}(t),\tilde{\bf r}_{0i}(t)\}$
and internal conformation variables $\{\tilde{\bf u}_i(z,t),{\bf u}_i(z,t)\}$
of flux lines, which are the true dynamical variables of the system, rather
than the averaged density $\rho({\bf r},z;t)=\langle\hat\rho({\bf r},z;t)\rangle$ which is a 
coarse-grained density with no detailed information on these conformation variables.
We shall first start by considering the case of a pure (disorder free) flux
liquid at equilibrium, which will serve as a starting point to our treatment of the
disordered case to be considered in Sec. \ref{Hartree} and needs therefore to
be as accurately specified and perfectly understood as possible.

\section{Equilibrium dynamics of pure flux
liquids}
\label{secEquilibrium}

In the free part ${\cal A}_{free}$ of the action, Eq. (\ref{Afree}), 
we rewrite the flux line trajectories $\{{\bf r}_i(z,t)\}$ and conjugate
fileds $\{\tilde{\bf r}_i(z,t)\}$ in the form:
\begin{subequations}
\begin{eqnarray}
{\bf r}_i(z,t) & = & {\bf r}_{0i}(t) + {\bf u}_i(z,t) \,,
\\
\tilde{\bf r}_i(z,t) & = &\tilde{\bf r}_{0i}(t) + \tilde{\bf u}_i(z,t) \,.
\end{eqnarray}
\end{subequations}
Using the fact that 
$\int dz\; {\bf u}_i(z,t) = \int dz\; \tilde{\bf u}_i(z,t)=0$, 
we easily obtain that the free part of the action 
${\cal A}_{free}$ can be written as the sum:
\begin{eqnarray}
{\cal A}_{free} = {\cal A}_{free}^{(0)} + {\cal A}_{free}^{(u)} \,,
\label{decAfree}
\end{eqnarray}
where 
\begin{equation}
{\cal A}_{free}^{(0)} =  \sum_{i=1}^N\int dt\;\Big\{
\frac{1}{2}(2L\gamma T)\tilde{\bf r}_{0i}^2(t) + 
i\tilde{\bf r}_{0i}(t)\,
L\gamma\partial_t{\bf r}_{0i}(t)
\Big\} \label{Afree0}
\end{equation}
depends only on c.m. variables, while
\begin{eqnarray}
{\cal A}_{free}^{(u)} & = & \sum_{i=1}^N\int dz\,dt\;\Big\{
\frac{1}{2}(2\gamma T)\tilde{\bf u}_{i}^2(z,t) + 
\nonumber\\
& + & i\tilde{\bf u}_{i}(z,t)\,\big[
\gamma\partial_t{\bf u}_{i}(z,t) - K\partial_z^2{\bf u}_i(z,t)
\big]\Big\}
\nonumber
\end{eqnarray}
is the free action for the internal modes of the flux lines. From Eq. (\ref{Afree0}),
it is not difficult to see that the c.m. mode of
flux lines is characterized by a friction coefficient  
$\gamma_0=L{\gamma}$, and hence that the diffusion constant $D_0$ of a free
flux line is inversely proportional to the thickness $L$ of the sample, as already
pointed out in refs. \cite{Marchetti-Nelson,Radzihovsky-Frey}:
\begin{equation}
D_0 = \frac{1}{L\gamma} \label{D0} \,.
\end{equation}
In the interaction part of the action ${\cal A}_{int}$,
we expand the interaction potential  
$V\big({\bf r}_i(z,t)-{\bf r}_j(z,t)\big)$ to linear order in the
displacement field:
\begin{eqnarray}
V\big({\bf r}_i(z,t) &-& {\bf r}_j(z,t)\big) =  
V\big({\bf r}_{0i}(t)-{\bf r}_{0j}(t)\big) +
\nonumber\\
&+& [{\bf u}_i(z,t) - {\bf u}_j(z,t)]\cdot\nabla 
V\big({\bf r}_{0i}(t)-{\bf r}_{0j}(t)\big).
\nonumber
\end{eqnarray}
The interaction part ${\cal A}_{int}$ can then be written in the form:
\begin{eqnarray}
{\cal A}_{int} = {\cal A}_{int}^{(0)} + {\cal A}_{int}^{(1)} \,,
\label{decAint}
\end{eqnarray}
with
\begin{subequations}
\begin{eqnarray}
{\cal A}_{int}^{(0)} & = & \sum_{i\neq j}\int dt\;
iL\tilde{\bf r}_{0i}(t)\cdot\nabla V\big({\bf r}_{0i}(t)-{\bf r}_{0j}(t)\big) \,,
\\
{\cal A}_{int}^{(1)} & = & \sum_{i\neq j}\int dz\,dt\;
i\tilde{u}_{i\alpha}(z,t)[u_{i\beta}(z,t) - u_{j\beta}(z,t)]\times 
\nonumber\\
&\times& \partial_\alpha\partial_\beta 
V\big({\bf r}_{0i}(t)-{\bf r}_{0j}(t)\big) \,.
\end{eqnarray}
\end{subequations}
Combining Eqs. (\ref{decAfree}) and (\ref{decAint}), we see that we can
already write the total action ${\cal A}$ in the form 
${\cal A}={\cal A}^{(0)} + {\cal A}^{(1)}$,
where 
\begin{eqnarray}
{\cal A}^{(0)} & = & {\cal A}_{free}^{(0)} + {\cal A}_{int}^{(0)} \,,
\nonumber\\
& = & \sum_{i=1}^N\int dt\;\Big\{
\frac{1}{2}(2L\gamma T)\tilde{\bf r}_{0i}^2(t) +
i\tilde{\bf r}_{0i}(t)\cdot\Big[
L\gamma\partial_t{\bf r}_{0i}(t) 
\nonumber\\
& + & L\nabla V\big({\bf r}_{0i}(t)-{\bf r}_{0j}(t)\big)\Big]
\label{A0}
\end{eqnarray}
depends exclusively on c.m. variables and can be thought of as the dynamical
MSR action of a liquid of hard rods of length $L$ interacting through the
potential $V_0({\bf r})=LV({\bf r})$. The effective action ${\cal A}^{(1)}$ is on the other hand given by: 
\begin{widetext}
\begin{eqnarray}
{\cal A}^{(1)} & = & \sum_{i=1}^N\int dz\,dt\;\Big\{
\frac{1}{2}(2\gamma T)\,\tilde{\bf u}_i(z,t) + i\tilde{\bf u}_i(z,t)\cdot\big[
\gamma\partial_t{\bf u}_i(z,t) - K\partial_z^2{\bf u}_i(z,t)
\big] + 
\nonumber\\
& + & i\tilde{u}_{i\alpha}(z,t)\Big[
\Big(\sum_{k(\neq i)}\partial_\alpha\partial_\beta 
V\big({\bf r}_{0i}(t) - {\bf r}_{0k}(t)\big)\Big)
\,\delta_{ij} - \sum_{j(\neq i)}\partial_\alpha\partial_\beta 
V\big({\bf r}_{0i}(t) - {\bf r}_{0j}(t)\big)\,
\Big]\;u_{j\beta}(z,t)
\Big\} \,,
\end{eqnarray}
\end{widetext}\noindent
and describes the internal fluctuations of the flux lines. As it stands,
however, ${\cal A}^{(1)}$ still contains c.m. dynamical variables. 
In order to obtain an effective action which depends only on the internal 
modes, we need to take the average of ${\cal A}^{(1)}$ over all 
configurations of the c.m. coordinates $\{{\bf r}_{0i}(t)\}$ which are
compatible with a liquid structure.
In the spirit of a cumulant expansion, \cite{Ettouhami1} we shall write
${\cal A}^{(u)} = \langle{\cal A}^{(1)}\rangle_0$,
where the average here is taken with statistical weight 
$\exp(-{\cal A}^{(0)})$, ${\cal A}^{(0)}$ being the dynamical action for the
c.m. mode, Eq. (\ref{A0})). Performing the above Gaussian average, we obtain:
\begin{eqnarray}
{\cal A}^{(u)} &=& \sum_{i=1}^N\int dz\,dt
\Big\{ \frac{1}{2}\,(2\gamma T)\tilde{\bf u}_i^2(z,t) 
\nonumber\\
&+& i\tilde{u}_{i\alpha}(z,t)\Big[
\Big(\big(\gamma\partial_t  - K\partial_z^2 \big)\delta_{\alpha\beta}+ 
\mu_{\alpha\beta}^{(i)}\Big)\,u_{i\beta}(z,t) 
\nonumber\\
&+& \sum_{j(\neq i)}\mu^{(ij)}_{\alpha\beta}u_{j\beta}(z,t)
\Big]\Big\} \,,
\end{eqnarray}
where we defined: \cite{Ettouhami1}
\begin{subequations}
\begin{eqnarray}
\mu^{(i)}_{\alpha\beta} & = & \Big\langle
\sum_{i\neq j}\partial_\alpha\partial_\beta V\big({\bf r}_{0i}(t) 
- {\bf r}_{0j}(t)\big)
\Big\rangle_0 \,,
\\
\mu^{(ij)}_{\alpha\beta} & = & - \Big\langle
\partial_\alpha\partial_\beta V\big({\bf r}_{0i}(t) - {\bf r}_{0j}(t)\big)
\Big\rangle_0 \,.
\end{eqnarray}
\end{subequations}
It is easy to see that
\begin{eqnarray}
\mu^{(i)}_{\alpha\beta} & = & \int d{\bf r}\,d{\bf r}'\;
\partial_\alpha\partial_\beta V\big({\bf r}-{\bf r}'\big) \times 
\nonumber\\
&\times&\big\langle\sum_{i\neq j}\delta({\bf r}-{\bf r}_{0i}(t))
\delta({\bf r}'-{\bf r}_{0j}(t))\big\rangle_0 \,,
\nonumber\\
& = & \rho^2 \int d{\bf r}\,d{\bf r}'\;
\partial_\alpha\partial_\beta V\big({\bf r}-{\bf r}'\big)  
\,g_0\big({\bf r}-{\bf r}'\big) \,,
\end{eqnarray}
where
\begin{eqnarray}
g_0({\bf r}-{\bf r}') = \frac{1}{\rho^2}
\big\langle\sum_{i\neq j}\delta({\bf r}-{\bf r}_{0i}(t))\delta({\bf r}'
-{\bf r}_{0j}(t))\big\rangle_0
\end{eqnarray}
is the (equal time) equilibrium pair distribution function of the c.m. mode of the
flux line liquid. Using the rotational symmetry of both $g_0({\bf r})$ 
and $V({\bf r})$, we obtain that 
$\mu^{(i)}_{\alpha\beta} = \mu\,\delta_{\alpha\beta}$ 
with \cite{Ettouhami1}:
\begin{eqnarray}
\mu = \frac{\rho}{d_\perp}\int d{\bf r}\; g_0(r)\nabla^2V(r) \,,
\label{eqmu}
\end{eqnarray}
which is the result (\ref{mu}) of Sec. \ref{Disorder};
and similarly that \cite{Ettouhami1}
\begin{eqnarray}
\mu^{(ij)}_{\alpha\beta} = -\frac{\mu\delta_{\alpha\beta}}{N-1} \,.
\end{eqnarray}
The above results lead to the following expression for the effective action of the
internal modes of flux lines:
\begin{eqnarray}
{\cal A}^{(u)} &=& \sum_{i=1}^N\int dz\,dt\;\Big\{ \frac{1}{2}
\,(2\gamma T)\tilde{\bf u}_i^2(z,t) 
\nonumber\\
&+&\sum_{j=1}^N i\tilde{u}_{i\alpha}(z,t)\Big[
\Big(\gamma\partial_t  - K\partial_z^2 + \frac{N}{N-1}\,\mu\,\Big)\delta_{ij}
\nonumber\\
&-& \frac{\mu}{N-1}\Big]\,{\bf u}_j(z,t)\,\Big\} \,.
\label{Aeff}
\end{eqnarray}
The above effective action can be written in Fourier space in the form:
\begin{eqnarray}
{\cal A}^{(u)}  & = &  \sum_{i,j}\!\sum_{n\neq 0}\int_{\omega}\!\Big[
\frac{1}{2}\tilde{\bf u}_i(q_n,\omega)
\tilde\Gamma_{ij}(q_n,\omega)\cdot\tilde{\bf u}_j(-q_n,-\omega) 
\nonumber\\
& + & i\tilde{\bf u}_{i}(q_n,\omega)
\Gamma_{ij}(q_n,\omega)\cdot{\bf u}_j(-q_n,-\omega)\,\Big] \,,
\label{AeffFourier}
\end{eqnarray}
where the shorthand $\int_\omega$ stands for
$\int_{-\infty}^\infty\frac{d\omega}{2\pi}$, and where the
dynamical kernels $\tilde\Gamma_{ij}(q_n,\omega)$ and
$\Gamma_{ij}(q_n,\omega)$ are given by:
\begin{subequations}
\begin{eqnarray}
\tilde\Gamma_{ij}(q_n,\omega) & = & 2\gamma TL\,\delta_{ij} \,,
\\
\Gamma_{ij}(q_n,\omega) & = &
L\Big[\big(-i\gamma\omega + K q_n^2 + \frac{N}{N-1}\mu\big)\delta_{ij}
\nonumber\\
& - &  \frac{\mu}{N-1}\Big] .
\label{kernelF}
\end{eqnarray}
\end{subequations}
In the thermodynamic limit $N\to\infty$, the kernel $\Gamma_{ij}(q_n,\omega)$
reduces to the diagonal form:
\begin{eqnarray}
\Gamma_{ij}(q_n,\omega) \simeq L\,
\big(-i\gamma\omega + K q_n^2 + \mu\big)\delta_{ij} \,.
\label{kernelFdiag}
\end{eqnarray}
We hence obtain in our perturbative approach that the internal modes of
different flux lines are effectively decoupled: in the thermodynamic
$N\to\infty$ limit, the effect of the interactions between vortices on their
{\em internal} fluctuations is entirely encoded in the $\mu$ term, which acts
as a quadratic confining potential ($\sim \frac{1}{2}\mu u_i^2$) 
for the {\em internal} modes of individual flux lines (in total agreement with the findings
of the static approach of ref.  \cite{Ettouhami1}).

Knowledge of the propagator $\Gamma_{ij}(q_n,\omega)$ allows us to 
find the reponse and correlation functions, $R_{ij}(q_n,\omega)$ and
$C_{ij}(q_n,\omega)$ respectively, which we define as follows:
\begin{subequations}
\begin{eqnarray}
R_{ij}(z-z',t-t') & = & \frac{\delta\langle 
u_{i\alpha}(z,t)\rangle}{\delta\zeta_{j\alpha}(z',t')} \,, 
\nonumber\\
& = & \frac{1}{d_\perp}\langle {\bf u}_i(z,t)\cdot 
i\tilde{\bf u}_j(z',t')\rangle \,,
\\
C_{ij}(z-z',t-t') & = & \frac{1}{d_\perp}
\langle {\bf u}_i(z,t)\cdot{\bf u}_j(z',t')\rangle \,.
\end{eqnarray}
\end{subequations}
Whithin the mean-field approach of this Section, 
and in the thermodynamic limit $N\to \infty$,
the above functions are both diagonal in the vortex
labels $i,\,j$. If we denote by $\Gamma(z,t)$, $R(z,t)$ and 
$C(z,t)$ the diagonal parts of the vertex, response and correlation functions 
respectively, in such a way that $\Gamma_{ij}(z,t)=\Gamma(z,t)\delta_{ij}$,
$R_{ij}(z-z',t-t')=R(z,t)\delta_{ij}$ and 
$C_{ij}(z-z',t-t')=C(z,t)\delta_{ij}$, then one can easily verify that \cite{SV,BMR,LDG}
\begin{subequations}
\begin{eqnarray}
R(q_n,\omega) & = & \frac{1}{\Gamma(q_n,\omega)} \,,
\label{RinvG}
\\
C(q_n,\omega) & = & \frac{ \tilde\Gamma(q_n,\omega)}{ |\Gamma(q_n,\omega)|^2} \,.
\label{CinvG}
\end{eqnarray}
\end{subequations}
In the thermodynamic ($N\to\infty$) limit, 
we obtain from Eqs. (\ref{kernelFdiag}) and (\ref{RinvG}) that
the response function $R(q_n,\omega)$ is given by: 
\begin{equation}
R(q_n,\omega)  \simeq \frac{1}{L(-i\gamma\omega + K q_n^2 + \mu)} \quad ,
\quad q_n\neq 0 \,.
\end{equation}
Performing a partial Fourier transform back to the variable $t$, we obtain:
\begin{eqnarray}
R(q_n,t) =  \frac{\theta(t)}{\gamma L}\,\mbox{e}^{-(\mu + K q_n^2)t/\gamma} \,,
\label{Rqnt}
\end{eqnarray}
where $\theta$ is Heaviside's unit step function. On the other hand, from
Eq. (\ref{CinvG}), we readily obtain for the 
correlation 
function $C(q_n,t)$ the following expression:
\begin{equation}
C(q_n,t) = \frac{T}{L(K q_n^2 +\mu)}\; \mbox{e}^{-(\mu + K q_n^2)\,|t|/\gamma}
\quad,\quad
q_n \neq 0 .
\label{Cqnt}
\end{equation}
It is easy to verify that the fluctuation-dissipation
relation 
\begin{eqnarray}
\theta(t)\partial_tC(q_n,t) = -TR(q_n,t) \quad,\quad q_n\neq 0
\label{FDT}
\end{eqnarray}
holds for the internal modes of flux lines, 
which indicates that these modes will eventually
reach thermal equilibrium at long enough times.
We however should emphasize that the above expressions of the response
and correlation functions are only valid for the internal 
modes of the flux lines. The c.m. mode of vortices, as described 
by the effective action (\ref{A0}), is still diffusive, although with a
diffusion constant $D$ which we expect to be reduced by the interactions to a value that is
smaller \cite{Leegwater} than the bare diffusion constant of free, noninteracting vortices
$D_0=1/(\gamma\,L)$ of Eq. (\ref{D0}). Indeed, from ref.  \cite{Leegwater}, one can estimate 
the value of the interacting diffusion constant $D$ in the absence of disorder to be of order:
\begin{eqnarray}
D \approx \frac{D_0}{1+2\pi\rho\xi^2 g_0(2a)}\,,
\end{eqnarray}
which is smaller than the bare diffusion constant $D_0$ for all values of applied magnetic fields 
smaller than the upper critical field $H_{c2}$.

Having derived the equilbrium dynamics of pure flux liquids, 
we now turn our attention to the more general case of
driven vortex liquids in the presence of a random pinning potential. 
It should be pointed out at this stage that the ``mass'' coefficient
$\mu$ given in Eq. (\ref{eqmu}), can be generalized to take into account 
large vortex distortions, as was done for the static case in
ref.  \cite{Ettouhami2}. This generalization is performed within a dynamic
Hartree approximation in Appendix B.

\section{Dynamics of driven flux-line liquids: perturbation theory}
\label{Hartree}

\subsection{Coarse-grained effective action for flux line dynamics}

Following refs.  \cite{BMR,SV,LDG}, we decompose the internal
modes of flux lines into short- and long-wavelength parts,
\begin{eqnarray}
{\bf u}_i(z,t) & = & {\bf u}_i^<(z,t) + {\bf u}_i^>(z,t) \,,
\label{ufastslow}
\end{eqnarray}
where (we here for convenience adopt a continuous notation for the 
$q_n$ summations):
\begin{subequations}
\begin{eqnarray}
{\bf u}_i^<(z,t) & = & \int_{q< \Lambda_<} {\bf u}_i(q_z,t)\,\mbox{e}^{iq_zz} \,,
\\
{\bf u}_i^>(z,t) & = & \int_{\Lambda_< < q < \Lambda} 
{\bf u}_i(q_z,t)\,\mbox{e}^{iq_zz} \,.
\end{eqnarray}
\end{subequations}
In the above equations, $\Lambda$ and $\Lambda_<$ are
high and low momentum cut-offs, respectively. The ultraviolet cutoff $\Lambda$ 
is given in terms of 
the superconducting coherence length $\xi_c$ along the direction of the
flux lines by $\Lambda=\pi/\xi_c$. Inserting the above decomposition,
Eq. (\ref{ufastslow}), and a similar decomposition
for the response field $\tilde{\bf u}_i(z,t)$,
into the dynamical action, we find after integrating out the short wavelength modes 
$\{ {\bf u}_i^>,\tilde{\bf u}_i^> \}$ that the long-wavelength effective action, to first
order perturbation theory, is given by:
\begin{equation}
{\cal A}_{eff} = {\cal A}_{pure} +
\langle {\cal A}_{dis}[\tilde{\bf u}^<+\tilde{\bf u}^>,{\bf u}^<
+{\bf u}^>]\rangle_{>} \,,
\label{defAeff}
\end{equation}
where
\begin{widetext}
\begin{eqnarray}
\langle {\cal A}_{dis}\rangle_{>} &=& 
\frac{1}{2}\sum_{i,j}\int\!dz\,dt\int\!dz'\,dt'\;
\tilde{r}_{i\alpha}^<(z,t)
\delta\tilde{\Gamma}_{ij}^{\alpha\beta}(z,t;z',t')
\tilde{r}_{j\beta}^<(z',t') +
\sum_{i} \int\!dz\,dt\;
i\tilde{\bf r}_{i\alpha}^<(z,t)
\delta\Gamma_{i\alpha}(z,t)  \,,
\label{avgaction}
\end{eqnarray}
where we defined
\begin{subequations}
\begin{eqnarray}
\delta\tilde{\Gamma}_{ij}^{\alpha\beta}(z,t;z',t') & = &
\int_{\bf q} iq_\alpha iq_\beta \Delta({\bf q},z-z')
\mbox{e}^{i{\bf q}\cdot[{\bf r}_i^<(z,t)-{\bf r}_j^<(z',t')]}
\mbox{e}^{-\frac{1}{2}q_\alpha q_\beta\phi_{ij}^{\alpha\beta}(z,t;z',t')} \,,
\label{delGamma_ij}
\\
\delta\Gamma_{i\alpha}(z,t;z',t') & = & (-i)
\sum_{j}\int dz'dt' R_{ij}(z,t;z',t')
\int_{\bf q} q_\alpha q^2\Delta({\bf q},z-z')
\mbox{e}^{i{\bf q}\cdot[{\bf r}_i^<(z,t)-{\bf r}_j^<(z',t')]}
\mbox{e}^{-\frac{1}{2}q_\alpha q_\beta\phi_{ij}^{\alpha\beta}(z,t;z',t')} \,,
\label{delGamma_i}
\end{eqnarray}
\end{subequations}
with $\phi_{ij}^{\alpha\beta}(z,t;z',t')=
\langle[u_{i\alpha}(z,t)-u_{i\alpha}(z',t')]
[u_{j\beta}(z,t)-u_{j\beta}(z',t')]\rangle$.
In perturbation theory, valid at large driving forces, it is convenient to
take the limit $\Lambda_< \to 0$, and 
use for the response and correlation functions the expressions 
(\ref{Rqnt})-(\ref{Cqnt}), which are spatially invariant and diagonal in vortex
indices, in which case expressions (\ref{delGamma_ij})-(\ref{delGamma_i})
are greatly simplified, and become:
\begin{subequations}
\begin{eqnarray}
\delta\tilde{\Gamma}_{ij}^{\alpha\beta}(z,t;z',t') & = &
\int_{\bf q} iq_\alpha iq_\beta \Delta({\bf q},z-z')
\mbox{e}^{i{\bf q}\cdot[{\bf r}_i^<(z,t)-{\bf r}_j^<(z',t')]}
\mbox{e}^{-\frac{q^2}{2d_\perp}\,\Phi(z-z',t-t')} \,,
\label{newdelGamma_ij}
\\
\delta\Gamma_{i}^{\alpha}(z,t;z',t') & = & \frac{-i}{d_\perp}\int dz'dt'\; R(z-z';t-t')
\int_{\bf q} iq_\alpha q^2
\Delta({\bf q},z-z')\mbox{e}^{i{\bf q}\cdot[{\bf r}_i^<(z,t)-{\bf r}_j^<(z',t')]}
\mbox{e}^{-\frac{q^2}{2d_\perp}\Phi(z-z';t-t')} \,.
\label{newdelgammai}
\end{eqnarray}
\end{subequations}
\end{widetext}
In the above expression, $\Phi(z,t)$ denotes the quantity
\begin{eqnarray}
\Phi(z,t) = \phi_0(t) + \phi(z,t)
\end{eqnarray}
where
$\phi_0(t)=\langle[{\bf r}_{0}(t) - {\bf r}_{0}(0)]^2\rangle_0$
is the relative displacement of the c.m. mode, while
\begin{eqnarray}
\phi(z,t) & = & \langle[{\bf u}_i(z,t)-{\bf u}_i(0,0)]^2\rangle \,,
\nonumber\\
& = & \frac{2Td_\perp}{L}\!\sum_{n=1}^\infty \frac{ 1 - \cos(q_nz)
\mbox{e}^{-(K q_n^2 + \mu)|t|/\gamma } }{K q_n^2 + \mu} \,.
\label{phi}
\end{eqnarray}
is the relative displacement of the internal mode ${\bf u}$ of a given flux line in the
vortex liquid.

Using Eqs. (\ref{defAeff}), (\ref{avgaction}) and (\ref{newdelgammai}), 
effective equations of motion for the c.m. and 
internal modes of flux lines can be derived in a standard way.  \cite{BMR,SV,LDG}
For the internal modes, we obtain
\begin{eqnarray}
\tilde{\gamma}_{\alpha\beta} \partial_t u_{i\beta}(z,t)  & = &  
(-\mu\delta_{\alpha\beta} + \tilde{K}_{\alpha\beta}\partial_z^2)u_{i\beta} +
\nonumber\\
& - &  \frac{1}{2} \lambda_{\alpha\beta\gamma}\partial_z u_{i\beta}
\partial_z u_{i\gamma} +   \zeta_i(z,t) \,,
\label{eqmotion}
\end{eqnarray}
where now, in addition to the usual (but renormalized) elastic tension term, new
non-linear Kardar-Parisi-Zhang (KPZ) terms have appeared.
The c.m. mode on the other hand obeys the following equation of motion:
\begin{equation}
\tilde{\gamma}_{\alpha\beta} \partial_t r_{0i\beta}(t)  = 
F_{\alpha} - F_{fr,\alpha} \,,
\end{equation}
where the friction force ${\bf F}_{fr}$ arises from the additional drag experienced
by the vortex liquid as a result of the presence of the random pinning
potential. In the following subsection, we outline the main steps of the
coarse graining procedure which leads to the above equation of motion, before
moving on in Sec. \ref{DynStrucFac} to calculating the dynamic structure
factor of the flux line liquid in presence of disorder.

\subsection{Derivation of renormalized quantities}

\subsubsection{Friction force}

The first order correction to the friction force is extracted from the
$\sim {\cal O}(\tilde{\bf r}_{0i})$ terms of the effective action of Eq. (\ref{defAeff}):
\begin{eqnarray}
{\cal A}_{eff}[\tilde{\bf r}_{0i}] & = & \frac{1}{d_\perp}\sum_i\int\!dt\; 
L\tilde{\bf r}_{0i\alpha}(t) 
\int\!dz\int_{\bf q} q_{\alpha}q^2
\Delta({\bf q},z) 
\nonumber\\
&\times & R(z,t)\mbox{e}^{i{\bf q}{\bf v}t-\frac{q^2}{2d_\perp}\Phi(z,t)}  \,.
\label{Afr}
\end{eqnarray}
This term is of the form:
\begin{eqnarray}
{\cal A}_{eff}[\tilde{\bf r}_{0i}] & = & \sum_i\int\!dt\; 
iL\tilde{\bf r}_{0i\alpha}(t)
F_{fr,\alpha} \,,
\end{eqnarray}
with the effective friction force:
\begin{equation}
F_{fr,\alpha}  \!=\! \int\!\!dzdt\!\!\int_{\bf q} \!\!
\frac{q_{\alpha}q^2}{d_\perp}
\Delta({\bf q},z) R(z,t)\sin({\bf q}\cdot{\bf v}t)
e^{-\frac{q^2}{2d_\perp}\Phi(z,t)}.
\end{equation}
This expression is identical to previously derived expressions \cite{Blatter-et-al} for the 
friction force on individual flux lines in presence of disorder, except that here the response 
and correlation functions to be used for an explicit evaluation of $F_{fr,\alpha}$ should be 
ones that are relevant to a flux liquid, e.g. Eqs. (\ref{Rqnt}) and (\ref{Cqnt}) respectively.

\subsubsection{Friction coefficient}

The disorder-correction to the friction coefficient is extracted from the effective
action as follows. In the expression (\ref{newdelgammai}) of the effective
kernel $\delta\Gamma_{i\alpha}$, we make use of the expansion
\begin{eqnarray}
\mbox{e}^{i{\bf q}\cdot[{\bf u}_i(z,t)-{\bf u}_j(z',t')]} &\simeq&
1 + iq_{\alpha}[u_{i\alpha}(z,t)-u_{j\alpha}(z',t')] 
\nonumber\\
& - & \frac{1}{2}q_\alpha q_\beta [u_{i\alpha}-u_{j\alpha}]
[u_{i\beta}-u_{j\beta}] + \ldots
\nonumber\\
\label{expansion}
\end{eqnarray}
We further shall assume that the disorder is weak, so that the internal
displacements of flux lines vary slowly on the
scale of the equilibrium kernel $\Gamma(z,t)$, i.e. on the scale of
$\sqrt{\mu/K}$. In this case, one can make use of the following gradient
expansion:
\begin{eqnarray}
u_{i\alpha}(z,t) &-& u_{j\alpha}(z',t') \simeq (t-t')\partial_t u_{i\alpha}(z,t) +
\nonumber\\
& + & (z-z')\partial_z u_{i\alpha} - 
\frac{1}{2}(z-z')^2\partial_z^2 u_{i\alpha} \,.
\label{taylor}
\end{eqnarray} 
Using both expansions (\ref{expansion}) and (\ref{taylor}) into
Eq. (\ref{newdelgammai}), we obtain that the effective action ${\cal A}_{eff}$
contains a term of the form:
\begin{eqnarray}
{\cal A}_{eff} = \sum_i\int dzdt \; i\tilde{r}^<_{i\alpha}(z,t)
[\delta\gamma_{\alpha\beta}\partial_t u_{i\beta}(z,t)] \,,
\end{eqnarray}
and hence that the effective friction coefficient
$\tilde{\gamma}_{\alpha\beta}$ in Eq. (\ref{eqmotion}) is given by:
\begin{equation}
\tilde{\gamma}_{\alpha\beta} = \gamma\delta_{\alpha\beta} +
\delta\gamma_{\alpha\beta} \,,
\end{equation}
with
\begin{equation}
\delta\gamma_{\alpha\beta}  =  \int dzdt\int_{\bf q} 
\frac{q_\alpha q_\beta q^2}{d_\perp}
\Delta({\bf q},z)\, tR(z,t)
\mbox{e}^{i{\bf q}\cdot{\bf v}t-\frac{q^2}{2d_\perp}\Phi(z,t)}.
\end{equation}
Note that in the absence of drive ($v=0$), $\tilde\gamma_{\alpha\beta}$
is isotropic, $\tilde\gamma_{\alpha\beta} =
(\gamma+\delta\gamma)\delta_{\alpha\beta}$, where now:
\begin{equation}
\delta\gamma  =  \frac{1}{d^2_\perp}\int dzdt\int_{\bf q} q^4
\Delta({\bf q},z)\, tR(z,t)
\mbox{e}^{-\frac{q^2}{2d_\perp}\Phi(z,t)}.
\end{equation}

\subsubsection{Elastic dispersion}

It also follows from Eqs.(\ref{expansion})-(\ref{taylor}) that the elastic
coefficients appearing in Eq. (\ref{eqmotion}) can
be written in the form 
\begin{eqnarray}
\tilde{K}_{\alpha\beta} = K\delta_{\alpha\beta} + \delta K_{\alpha\beta} \,,
\end{eqnarray}
where the disorder-dependent correction $\delta K_{\alpha\beta}$ is given by:
\begin{eqnarray}
\delta K_{\alpha\beta} & = & \frac{1}{2d_\perp}\int dzdt\int_{\bf q} q_\alpha q_\beta 
q^2z^2\Delta({\bf q},z) R(z,t) \times\nonumber\\
&\times& \mbox{e}^{i{\bf q}\cdot{\bf v}t -\frac{q^2}{2d_\perp}\Phi(z,t)} \,.
\end{eqnarray}
This equation can be rewritten, using the fluctuation-dissipation theorem, 
Eq. (\ref{FDT}), in the form
\begin{eqnarray}
\delta K_{\alpha\beta} & = & -\int dz\int_{\bf q} q_\alpha q_\beta 
z^2\Delta({\bf q},z) \times\nonumber\\
&\times& \int_0^{\infty} dt\;
e^{i{\bf q}\cdot{\bf v}t  -\frac{q^2}{2d_\perp}\phi_0(z,t)}
\partial_t e^{ -\frac{q^2}{2d_\perp}\phi(z,t)} \,.
\end{eqnarray}
In the static ($v=0$) limit, the above correction to the tilt modulus becomes isotropic,
$\delta K_{\alpha\beta}=\delta K \delta_{\alpha\beta}$, with:
\begin{equation}
\delta K  =  \frac{1}{d_\perp T}\int dz\int_{\bf q} q^2    
z^2\Delta({\bf q},z) e^{ -\frac{q^2}{2d_\perp}\Phi(z,0)} + {\cal O}(1/L)\,,
\end{equation}
which is the result (\ref{K_R}) that we obtained in Sec. \ref{GVM} within
a static replica approach.

The Taylor expansion (\ref{taylor}) also yields convective terms of
the form $\chi_{\alpha\beta}\partial_z u_{i\beta}(z,t)$ on the right hand side of
Eq. (\ref{eqmotion}). It is easy to see, however, that the coefficients of these
convective terms,
\begin{eqnarray}
\chi_{\alpha\beta} & = & \frac{1}{2d_\perp}\int dzdt\int_{\bf q} q_\alpha q_\beta q^2
z\Delta({\bf q},z) R(z,t) \times
\nonumber\\
&\times& \mbox{e}^{i{\bf q}\cdot{\bf v}t}\mbox{e}^{-\frac{q^2}{2d_\perp}\Phi(z,t)},
\end{eqnarray}
vanish identically by virtue of the fact that the integrand is odd in the
integration variable $z$ (provided that the disorder correlator is even in $z$, 
i.e. $\Delta({\bf r},-z)=\Delta({\bf r},z)$).

\subsubsection{The KPZ nonlinearity}

The effective action (\ref{defAeff}) contains an additional contribution
of the form
\begin{eqnarray}
{\cal A}_{eff}[\tilde{u}u^2] & = & \sum_i\int dzdt\, i\tilde{r}^<_{i\alpha}(z,t)
\times
\nonumber\\
&\times& \big\{
-\frac{1}{2}\lambda_{\alpha\gamma\rho}[\partial_zu_\delta(z,t)]
[\partial_zu_\rho(z,t)]
\big\},
\nonumber\\
\end{eqnarray}
with  
\begin{eqnarray}
\lambda_{\alpha\beta\gamma} & = & \frac{-i}{d_\perp}\int dzdt\int_{\bf q} 
q_\alpha q_\beta q_\gamma q^2 z^2\Delta({\bf q},z) R(z,t) \times
\nonumber\\
&\times& \mbox{e}^{i{\bf q}\cdot{\bf v}t}\mbox{e}^{-\frac{q^2}{2d_\perp}\Phi(z,t)}\,.
\end{eqnarray}
This means that disorder induces a KPZ nonlinearity in the driven state, much
as it does for driven vortex solids. In fact, the emergence of non-linear KPZ
terms in driven flux-line liquids has been predicted a long time ago 
within a macroscopic approach by Hwa \cite{Hwa},
who found that these terms affect the dynamics of the flux liquid on long length scales,
with the vortices forming a smooth, laminar phase at small drives, and a rough, turbulent
phase a large values of the applied force. We shall not study the effect of the KPZ terms
in any more detail here, and instead refer the reader to the above mentioned paper
for more details on this particular question.

\section{Dynamic structure factor of disordered flux line liquids}
\label{DynStrucFac}

We are now in a position to calculate the dynamic structure factor 
$S({\bf r},z;t)$ of our flux line liquid. By definition: 
\begin{eqnarray}
S({\bf r},z;t) = \big\langle
\hat\rho({\bf r},z;t)\hat\rho({\bf 0},0;0) 
\big\rangle \,,
\end{eqnarray}
where $\langle\cdots\rangle$ now stands for averaging over both c.m. and
internal conformation variables of vortices, and where 
space translational invariance of the flux-liquid has been assumed.
In what follows, it will be convenient to consider the partial Fourier transform
\begin{eqnarray}
S({\bf q},z;t) = \frac{1}{L_\perp^{d_\perp}}\big\langle
\hat\rho({\bf q},z;t)\hat\rho(-{\bf q},0;0) 
\big\rangle \,,
\label{Eq:def_S}
\end{eqnarray}
where $L_\perp$ is the size of the system in the plane perpendicular to flux lines.
Using the definition of the density operator at time $t$,
\begin{eqnarray}
\hat\rho({\bf r},z;t) = \sum_{i=1}^N \delta\big({\bf r}-{\bf r}_i(z,t)\big) \,,
\end{eqnarray}
we readily obtain that the Fourier transform $S({\bf q},z;t)$ is given by
\begin{eqnarray}
S({\bf q},z;t) \!=\! \frac{1}{L_\perp^{d_\perp}}\sum_{i=1}^N\sum_{j=1}^N\Big\langle
\mbox{e}^{-i{\bf q}\cdot[{\bf r}_i(z,t)-{\bf r}_j(0,0)]}
\Big\rangle \,.
\label{Sqzt}
\end{eqnarray}
We now separate the c.m. mode from the internal modes of 
the flux lines, and assume that the dynamical action ${\cal A}$ has 
been written in the decoupled form:
\begin{eqnarray}
{\cal A} = {\cal A}^{(0)} + {\cal A}^{(u)}
\end{eqnarray}
where ${\cal A}^{(0)}$ depends only on the c.m. variables $\{{\bf r}_{0i}(t)\}$, 
while ${\cal A}^{(u)}$ only depends on the internal modes $\{{\bf u}_i(z,t)\}$.
It then follows that the average on the right hand side of Eq. (\ref{Sqzt}) can be written in the form:
\begin{eqnarray}
\Big\langle
\mbox{e}^{-i{\bf q}\cdot[{\bf r}_i(z,t)-{\bf r}_j(z',t')]}\Big\rangle 
& = & \Big\langle \mbox{e}^{-i{\bf q}\cdot[{\bf r}_{0i}(t)-{\bf r}_{0j}(t')]}\Big\rangle_0
\nonumber\\
&\times& \mbox{e}^{-\frac{1}{2d_\perp}\,q^2\langle[{\bf u}_i(z,t)-{\bf u}_j(z',t')]^2\rangle_u} \,.
\nonumber\\
\end{eqnarray}
In the above expressions, $\langle\cdots\rangle_0$ and $\langle\cdots\rangle_u$ denote averages with statistical 
weights $\exp(-{\cal A}^{(0)})$ and $\exp(-{\cal A}^{(u)})$, respectively.
Now, in the approaches of Secs. \ref{SecMSR} and
\ref{Hartree}, the internal modes of different flux lines are 
decoupled, which implies that, for $i\neq j$,
\begin{eqnarray}
\langle[{\bf u}_i(z,t)-{\bf u}_j(z',t')]^2\rangle
= 2\langle u^2(z,t)\rangle \,,
\end{eqnarray}
and so the expression (\ref{Sqzt}) of $S({\bf q},q_z;t)$ becomes
\begin{eqnarray}
S({\bf q},z;t) & = & \frac{1}{L_\perp^{d_\perp}}\Big\{
\sum_{i=1}^N\Big\langle
\mbox{e}^{-i{\bf q}\cdot[{\bf r}_{0i}(t)-{\bf r}_{0i}(0)]}
\Big\rangle_0
\mbox{e}^{-\frac{q^2}{d_\perp}\phi(z,t)} 
\nonumber\\
& + &\sum_{i=1}^N\sum_{j\neq i}\Big\langle
\mbox{e}^{-i{\bf q}\cdot[{\bf r}_{0i}(t)-{\bf r}_{0j}(0)]}
\Big\rangle_0
\mbox{e}^{-\frac{q^2}{d_\perp}\langle u^2(z,t)\rangle} 
\Big\} .
\nonumber\\
\label{Sqzt2}
\end{eqnarray}
Given that all vortices in the flux liquid are equivalent
to each other in our mean field approach, we see that the first term on the
right hand side of Eq. (\ref{Sqzt2}) involves the sum of $N$ identical terms,
which we can simply write as $NF({\bf q},t)$, with: \cite{Remark}
\begin{eqnarray}
F({\bf q},t) & = & \frac{1}{N}\sum_{i=1}^N\Big\langle
\mbox{e}^{i{\bf q}\cdot[{\bf r}_{0i}(t)-{\bf r}_{0i}(0)]}
\Big\rangle_0 \,,
\nonumber\\
& \simeq & \mbox{e}^{-\frac{q^2}{2d_\perp}\phi_0(t)} \,.
\label{secondline}
\end{eqnarray}
On the other hand, it is easy to verify that
\begin{eqnarray}
\sum_{i\neq j}\Big\langle
\mbox{e}^{-i{\bf q}\cdot[{\bf r}_{0i}(t)-{\bf r}_{0j}(t')]}\Big\rangle_0 
= L_\perp^{d_\perp}\,\rho^2g_0({\bf q},t-t') \,,
\end{eqnarray}
where $g_0({\bf q};t-t')$ is the Fourier transform of 
the time dependent version of the pair distribution function of standard liquid state 
theory, which, in real space, is given by:
\begin{equation}
g_0({\bf r}-{\bf r}',t-t') \!=\! \frac{1}{\rho^2}\sum_{n\neq m}\big\langle
\delta\big({\bf r}\!-\!{\bf r}_{0n}(t)\big)\delta\big({\bf r}'\!-\!{\bf r}_{0m}(t')\big)
\big\rangle_0.
\end{equation}
Using the fact that $N=L_\perp^{d_\perp}\rho$, we 
finally obtain
\begin{equation}
S({\bf q},z;t)  =  \rho\mbox{e}^{-\frac{1}{2d_\perp}\,q^2\Phi(t)}
+\rho^2g_0({\bf q},t-t')\;\mbox{e}^{-\frac{1}{d_\perp}\,q^2\langle u^2(z,t)\rangle} 
\label{interm-S-1}
\end{equation}
A principal difficulty with the expression above is to find a good approximation for the 
time-dependent pair 
correlation function $g({\bf q},t-t')$. The simplest such approximation is the so-called 
``convolution approximation'', which was introduced many years ago by Vineyard, \cite{Vineyard} 
and which consists in writing for $g_0({\bf q},t)$ the following expression:
\begin{eqnarray}
g_0({\bf q},t) = g_0({\bf q})\,F({\bf q},t) \,.
\end{eqnarray}
Here $g_0({\bf q})=g_0({\bf q},t=0)$ is the ordinary (equal time) 
pair distribution function of static liquid 
state theory, and $F({\bf q},t)$ is the quantity defined in Eq. (\ref{secondline}).
Using this approximation into Eq. (\ref{interm-S-1}) above leads to the following result:
\begin{equation}
S({\bf q},z;t) \! =\!  \rho\mbox{e}^{-\frac{q^2}{2d_\perp}\phi_0(t)}\Big\{
\rho\,g_0({\bf q})\mbox{e}^{-\frac{q^2}{d_\perp}\langle u^2\rangle} 
+ \mbox{e}^{-\frac{q^2}{2d_\perp}\phi(z,t)}   
\Big\},
\label{Sqztfinal}
\end{equation}
where we used Eq. (\ref{secondline}) to express $F({\bf q},t)$ in terms of $\phi_0(t)$.
In the case where the c.m. mode of the flux lines obeys a simple diffusion law of the form:
\begin{eqnarray}
\phi_0(t) = 2d_\perp TD\,|t| \,,
\end{eqnarray}
we obtain the following expression for the structure factor $S({\bf q},z;t)$
\begin{equation}
S({\bf q},z;t)   \!=\!   \rho\mbox{e}^{-DT|t|q^2}\Big[
\rho\,g_0({\bf q})\mbox{e}^{-\frac{1}{d_\perp}\,q^2\langle u^2\rangle} +
\mbox{e}^{-\frac{q^2}{2d_\perp}\phi(z,t)}
\Big] \,.
\label{Sqztnew}
\end{equation}
In the following, last Section, we shall discuss some limiting cases, and compare our 
theoretical prediction for the static structure factor to experimental results.

%%%%%%%%%%%%%%%%

\section{Discussion and Conclusions}
\label{Conclusion}

We now discuss the meaning and phsical implications of our results. 
We shall start by addressing the nature of the unusual 
massive phonon mode of the internal fluctuations of flux-lines, 
which seems to violate translational invariance. 
The first occurrence of this massive mode is in Eq. (\ref{Hpure1}), and it 
is very easy to verify that this equation is translationally invariant 
(even though this is not obvious from the way it is written).
In Eq. (\ref{result-Heff}), the off-diagonal $\mu_{ij}$ terms, which are necessary to 
maintain translational invariance, were discarded for simplicity.
Keeping these off-diagonal terms only leads to corrections of order 1/N to 
the elastic propagator (as was shown in detail in ref.  \cite{Ettouhami1})
which vanish in the thermodynamic limit. In fact, 
even though a massive term may seem unusual, it is 
very well known that massive phonon modes {\em do} appear in ordinary crystal 
lattices if the lattice has a basis (these are the so-called \cite{Ashcroft} ``optical 
phonons"). In the case of flux liquids, the internal modes of the flux 
lines are the analogues of atoms belonging to the basis, and the $\mu$ terms 
may be thought of as the analogues of ``optical phonons" in crystals.

We now turn our attention to the static structure factor $S({\bf r},z)$ of
the flux line liquid, which is defined as:
\begin{eqnarray}
S({\bf r},z) = \langle\hat\rho({\bf r},z)\hat\rho({\bf 0},0)\rangle \; .
\end{eqnarray}
In ref. \cite{Ettouhami2}, we have shown that for an effective Hamiltonian
$H_u$ of the general form:
\begin{eqnarray}
H_u = \frac{1}{2}\sum_{i=1}^N\sum_{n\neq 0}
G^{-1}(q_n)|{\bf u}_i(q_n)|^2 \, ,
\label{genHu}
\end{eqnarray}
the structure factor is given by:
\begin{eqnarray}
S({\bf q},z) = \rho^2 g_0({\bf q}) \,\mbox{e}^{-q^2\langle u^2\rangle/d_\perp} +
\rho\mbox{e}^{-\frac{1}{2d_\perp}q^2\phi(z)} \,,
\label{generalS}
\end{eqnarray}
where the mean square relative displacement $\phi(z)$ has already been evaluated in
Eq. (\ref{phiofz}),
and where the mean projected area $\langle u^2\rangle$ of a given flux line is
given in terms of the elastic propagator $G$ by:
\begin{eqnarray}
\langle u^2\rangle = d_\perp T \sum_{n\neq 0} G(q_n) \,.
\end{eqnarray}
In the present case of a disordered vortex liquid with the inverse elastic
propagator of Eq. (\ref{genericform}), the above expression for $\langle u^2\rangle$ yields:
\begin{equation}
\langle u^2\rangle = \frac{d_\perp T}{2\sqrt{\mu_R K_R} } .
\label{result-u2}
\end{equation}
We thus obtain for the structure factor the following expression:
\begin{eqnarray}
S({\bf q},z) &=& \rho^2 g_0({\bf q})
\,\exp\Big(-\frac{Tq^2}{2\sqrt{{\mu_R}K_R}}\Big) 
\nonumber\\
&+&\rho\exp\Big\{-\frac{Tq^2}{2\sqrt{{\mu_R}K_R}}\big[ 1
- \mbox{e}^{-|z|\sqrt{\mu_R/K_R}}\big]\Big\}.
\nonumber\\
\label{ourS}
\end{eqnarray}
The above result for the structure factor is very different from the result
obtained by previous authors in
the hydrodynamic formulations of refs.  \cite{Nelson-Seung,Nelson-LeDoussal,Tauber-Nelson,Radzihovsky-Frey,Benetatos-Marchetti}
which is given by:
\begin{equation}
S({\bf q},q_z)  =  \frac{\rho Tq^2/{K}}{q_z^2+\varepsilon^2({\bf q})/T^2} +
\Delta({\bf q},q_z)\Big[
\frac{\rho q^2/{K}}{q_z^2+\varepsilon^2({\bf q})/T^2}
\Big]^2 \; ,
\label{strucfac}
\end{equation}
where the excitation spectrum has the usual 
bosonic \cite{Bogoliubov} form:
\begin{eqnarray}
\frac{\varepsilon({\bf q})}{T} = \Big[
\Big(\frac{Tq^2}{2K}\Big)^2 + \frac{\rho V(q)q^2}{K}
\Big]^{1/2} \,.
\label{spectrum}
\end{eqnarray}
In ref. \cite{Ettouhami2} we made a number of comments on the structure
factor of interacting, but otherwise disorder-free flux line liquids derived in
the boson mapping approach, Eq. (\ref{strucfac}), and revealed that it had a number
of quite disturbing inconsistencies. We here make
the similar observation that while the form (\ref{generalS}) follows from
the very general assumption that a decomposition of the form (\ref{decomp}) (with
an arbitrary elastic propagator $G(q_n)$ in Eq. (\ref{defH1})), can be written for
the Hamiltonian of the vortex liquid, the structure factor of Eq. (\ref{strucfac}) 
does not correspond to {\em any} choice of elastic propagator $G(q_n)$, and cannot 
be derived from a microscopic approach like ours. In Appendix \ref{App:Hydrodynamics}
we show that an expression for the structure factor that is similar 
to Eq. (\ref{strucfac}) (with an additive contribution proportional to disorder) can be
derived for the correlations of the density of the {\em center of mass} coordinates of flux lines. 
One may therefore speculate that the discrepancy between our results and those
of refs.  \cite{Nelson-Seung,Nelson-LeDoussal,Tauber-Nelson,Benetatos-Marchetti}
are due to the fact that in these previous studies the 
non-diffusive internal modes of vortices were treated on equal footing with the
c.m. mode (a diffusive mode for which hydrodynamics is naturally expected to be valid).
A more detailed discussion of the formulation of Gaussian hydrodynamics
of continuous media and of the shortcomings of previous attempts to formulate the hydrodynamics
of flux-line liquids can be found in Appendix \ref{App:Hydrodynamics}.

% Major change:
%--------------
%It is in fact very surprising that the structure factor
%of Eq. (\ref{strucfac}) contains a {\em separate} additive contribution coming from
%the disorder, when known results for the structure factor of the disordered flux
%line {\em lattice}  \cite{GLD} do {\em not} contain such an additional
%contribution. In other words, using the boson mapping to find the
%structure factor of a disordered vortex lattice would give an expression 
%similar to Eq. (\ref{strucfac}) for
%the structure factor, which we now know is an incorrect result for the structure
%factor of a pinned flux-line lattice.
%In fact, our results can be seen as the logical 
%extension of the replica variational treatments of flux line lattices  \cite{GLD} 
%which predict a finite renormalization of the elastic coefficients (in case the 
%Gaussian disorder does not have ultralocal delta-function-like correlations, which is the case 
%we consider in the present study), but where 
%disorder does {\em not} lead to any {\em additive} 
%term in the structure factor (in agreement 
%also with functional renomalization group results). 
%

We now consider some limiting cases. For a ``perfect gas'' of noninteracting
flux lines, $g_0(r)=1$, and Eq. (\ref{Sqztfinal}) reduces to:
\begin{equation}
S({\bf q},z;t)  =   (2\pi)^2\rho^2\,\delta({\bf q}) + \rho\,\mbox{e}^{-D_0T|t|\,q^2}
\mbox{e}^{-\frac{q^2}{2d_\perp}\phi(z,t)} \,,
\label{Sfree1}   
\end{equation}
with the diffusion constant of the noninteracting c.m. mode $D_0=1/(\gamma L)$.
For noninteracting flux lines, the correlation function $\phi(z,t)$ is given by:
\begin{eqnarray}
\phi(z,t) & = & \frac{2Td_\perp}{L}\sum_{n=1}^\infty \frac{1}{K q_n^2}\big[
1 - \cos(q_nz)\mbox{e}^{-K q_n^2\frac{|t|}{\gamma}}
\big] \,.
\label{phifree}
\end{eqnarray}
At times shorter than the characteristic Rouse time $t_{Rouse}=\gamma
L^2/K$, the sum can be transformed into an 
integral, with the result  \cite{Radzihovsky-Frey}:
\begin{eqnarray}
\phi(t) = \frac{2Td_\perp}{K}\,|z|\,f\Big(\frac{K|t|}{\gamma z^2}\Big) \,,
\end{eqnarray}
where $f$ is the function given by (here $\Gamma(a,x)$ is the incomplete
gamma function  \cite{Abramowitz}):
\begin{eqnarray}
f(u) & = & \frac{1}{\pi}\int_0^\infty \frac{dx}{x^2} \;[ 1-
\mbox{e}^{-ux^2}\cos{x}] \,,
\nonumber\\
& = & \frac{1}{2} +
\frac{1}{4\sqrt{\pi}}\Gamma\Big(-\frac{1}{2},\frac{1}{4u}\Big) \,,
\end{eqnarray}
with the limiting behavior $f(u)\simeq 1/2$ for $u\to 0$ and $f(u)\simeq
\sqrt{u}/\pi$ for $u\gg 1$.
On the other hand, at long enough times, $t>t_{Rouse}$, the sum in
Eq.(\ref{phifree}) is rapidly converging to the 
limiting value:
\begin{eqnarray}
\phi(t\gg t_{Rouse}) & = &
\frac{2Td_\perp}{LK}\,\sum_{n=1}^\infty\frac{1}{q_n^2} \,,
\nonumber\\
& = & \frac{LTd_\perp}{12K} \,.
\end{eqnarray}
It thus follows that, at long times ($t>t_{Rouse}$), the structure factor of
noninteracting flux lines can be written in the form:
\begin{eqnarray}
S({\bf q},z;t) & = &  \rho\,\mbox{e}^{-D_0T|t|\,q^2}\,\big[ 1 +
(2\pi)^2\rho\,\delta({\bf q})\big]
\mbox{e}^{-\frac{LTq^2}{24K}} 
\nonumber\\
& \approx & (2\pi)^2\rho^2\delta({\bf q})\,.
\label{Sfree2}
\end{eqnarray}  
Let us now consider the other limiting case of a liquid of infinitely rigid
flux lines. If we formally let $K\to\infty$ in our equations, it is not
difficult to verify that $\langle u^2\rangle=0$ and $\phi(z,t)=\langle[{\bf
u}_i(z,t)-{\bf u}_i(0,0)]^2\rangle = 0$, and thus  Eq. (\ref{Sqztfinal}) 
reduces to the appropriate expression for a liquid of point particles, in this
case the liquid formed by the centers of mass of the different vortices:
\begin{eqnarray}
S({\bf q},z;t)  =  \rho\;\mbox{e}^{-\frac{1}{2d_\perp}\,q^2\phi_0(t)}\;
[ 1 + \rho\,g_0({\bf q}) ] \,.
\label{Srigid}
\end{eqnarray}
Note that the $z$ dependence has dropped from this last equation.
Now, since , by definition, $\phi_0(0)=0$, we see that the equal time
structure factors $S(q,z;t=0)$ of Eqs. (\ref{Sqztfinal}), (\ref{Sqztnew}) 
and (\ref{Srigid}) reduce to the corresponding quantities
obtained in refs.  \cite{Ettouhami1,Ettouhami2}. 
In particular, for a liquid of rigid flux lines, Eq. (\ref{Srigid})
reduces to the correct expression of the static structure factor of standard
liquid state theory,
\begin{eqnarray}
S({\bf q},z)=  \rho\,[ 1 + \rho\,g_0({\bf q}) ] \,,
\label{Srigidstatic}
\end{eqnarray}
an expression which cannot be reproduced using boson mapping and other 
hydrodynamic methods.

We now turn our attention to the interacting structure factor of Eq. (\ref{Sqztfinal}). 
By contrast to the case of non-interacting flux-lines treated above, we see here that
due to the presence of the confining term $\mu$, the relaxation of the
internal modes is extremely fast: on time scales larger than the
characteristic time $t_\mu=\gamma/\mu$, the correlation function
$\phi(z,t)$ of Eq. (\ref{phi}) reaches its limiting value:
\begin{eqnarray}
\phi(z,t>t_\mu) & = & \frac{2Td_\perp}{\pi}\int_0^\infty 
\frac{dq}{K q^2 + \mu} \,,
\nonumber\\
& = &\frac{Td_\perp}{\sqrt{K\mu}} \,,
\end{eqnarray}
which implies that the low frequency behavior of the correlation function
$S(q,z;\omega)$ is given by:
\begin{eqnarray}
S({\bf q},z;\omega)   =  \frac{2\rho {DT}q^2}{\omega^2 + {D T}^2q^4}\;
\big[ 1 + \rho\,g_0({\bf q}) \big]\,\mbox{e}^{-\frac{Tq^2}{\sqrt{K\mu}}} \,,
\label{Smu}
\end{eqnarray}
By taking the limit $K\to\infty$, we again see that the low frequency behavior of the 
dynamic structure factor of an interacting liquid of hard rods is identical
to the $\omega$ behavior of the corresponding noninteracting system, provided 
the bare diffusion constant $D_0$ is replaced by the renormalized quantity
$D$, and that the static structure of the liquid is taken into account
through the factor $[1+\rho g_0({\bf q})]$. For finite $K$, the only effect of the internal
fluctuations of the vortices on the structure factor on long time scales is
to introduce the additional ``Debye-Waller" factor 
$\exp(-Tq^2/\sqrt{K\mu})$. In the limit of noninteracting flux
lines, where $\mu\to 0$ and $D\to D_0$, the exponent 
$(Tq^2/\sqrt{K\mu})$ in this last factor goes to its upper bound
$(LTq^2/24K)$, the pair distribution function 
$g_0({\bf q})\to \delta({\bf q})$, and we recover the result
(\ref{Sfree2}) of an ideal gas of vortices.

\begin{figure}[th]
\includegraphics[width=8.09cm, height=5cm]{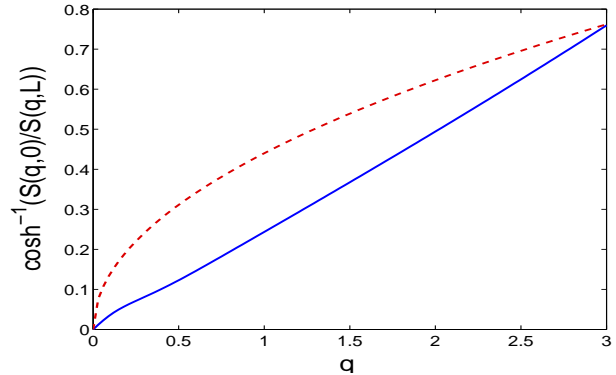}
\caption[]{(Color online) Solid line: Plot of $\cosh^{-1}\big[S(q,0)/S(q,L)\big]$ vs. $q$, using our result for the static
structure factor, Eq. (\ref{ourS}). Dashed line: approximate experimental curve, from ref.  \cite{Yoon}.
As in this last reference, the wavevector $q$ is measured in units of $q_{BZ}=(4\pi/\sqrt{3}a)$.
}\label{fig1}
\end{figure}

Going beyond the above limiting cases, we here would like to comment on the experiments of 
Yao {\em et al.} \cite{Yao} and Yoon 
{\em et al.} \cite{Yoon} who measured the structure factor of the vortex liquid in 
Bi$_2$Sr$_2$CaCu$_2$O$_8$ (BSCCO), and attempt a quantitative fit of experimental data using our 
theoretical prediction. For the extremely dilute vortex liquid studied in these experiments, the 
tilt modulus $K$ is given by the single vortex value 
$K\approx \varepsilon_0$, which has the numerical value \cite{Blatter-et-al} 
$\varepsilon_0(K/\AA)=1.964\times 10^8/[\lambda(\AA)]^2$.
Fig. \ref{fig1} shows a plot of $\cosh^{-1}\big[S(q,z=0)/S(q,z=L)\big]$ using the 
experimental parameters of ref.  \cite{Yoon}, 
namely:
\begin{subequations}
\begin{eqnarray}
a &\simeq& 1.5 \times 10^4 \AA \,,
\\
\lambda & \simeq & 0.62\times 10^4 \AA \,,
\\
L & \simeq & 0.2 \,\mbox{mm} \,,
\\
T & = & 80\,\mbox{K}\,,
\end{eqnarray}
\end{subequations}
and with the fit parameters $\alpha=1$ and $\eta=0.2$. Comparing our plot with the experimental 
curves (Fig. 2 of ref.  \cite{Yoon}), we see that our mean-field approach is 
able to produce a reasonably good qualitative fit of the data, which is quite surprising, given 
the rather simplified form of our model Hamiltonian, Eq. (\ref{def-H}), and of our analytic 
ansatz for the pair distribution function $g_0(r)$, Eq. (\ref{g0}).

It is worth nothing at this point that the confining coefficient $\mu$
can be reproduced (up to numerical factors of order unity) by taking the short wavelength limit
$q\to q_{BZ}$ (short wavelength fluctuations being the dominant ones in a liquid and 
$q_{BZ}=(4\pi/\sqrt{3}a)$ being the wavevector at the Brillouin zone boundary of a solid at the same
density) of the compression modulus $c_{11}(q)=B^2/4\pi(1+\lambda^2q^2)$
of usual elasticity theory. \cite{Brandt} That we are able to fit the experimental
data with a value of the compression modulus that qualitatively agrees with elasticity
theory is rather reassuring, and strongly supports our claim that an approach based on
conformation variables of vortices is more adequate to describe flux-line liquids than boson
mapping methods which by contrast, when used to fit the data of Yoon {\em et al.} \cite{Yoon},
give a result for $c_{11}$ which is smaller than the theoretical prediction by three orders of
magnitude. \cite{Yoon,CManalysis}

\noindent

The failure of hydrodynamic approaches to describe experimental data in a way that is consistent 
with standard elasticity theory is further indication of the importance of separating internal 
modes and c.m. variables of the vortices in the liquid state. Indeed, there is a very important qualitative distinction
between the c.m. mode on one hand and the internal modes on the other, for while the c.m. mode is diffusive,
internal modes of continuous media are non-diffusive (due to the elastic restoring forces, and this 
regardless of whether the confining mass $\mu$ is zero or not). The results of the present study outline
the importance of separating the non-diffusive modes from the diffusive ones 
that can be studied using hydrodynamic treatments.

%Major change:
%-------------
%Indeed, the basic idea of hydrodynamic approaches 
%is to write the free energy in terms of coarse-grained variables using symmetry considerations. 
%While this approach is quite successful in describing the long-wavelength behavior 
%of ordinary classical liquids of point particles, special care must be exercised when 
%dealing with continuous media, such as flux lines, 
%if one is to achieve quantitative agreement with experiments.

Another quantity which deserves attention is the friction force 
experienced by the flux liquid driven in presence of disorder.
For a flux liquid in presence of point disorder with correlations
$\Delta({\bf r},z) = \Delta_0\exp(-r^2/2\xi^2)\delta(z)/{2\pi\xi^2}$
we obtain, in the large $v$ limit and in $d=3$ dimensions:
\begin{eqnarray}
F_{fr} \sim
\Delta_0\frac{\sqrt{\gamma/K}}{\tilde\xi^{9/2}}\frac{1}{\sqrt{v}}\,,
\end{eqnarray}
where factors of order unity have been dropped, and where we defined
the length $\tilde\xi$ such that
$\tilde{\xi}^2 = \xi^2 + \langle u^2\rangle$. We see that the friction force vanishes
at large drives in agreement with previous predictions for vortex lattices and liquids in the 
plastic regime near the melting temperature. \cite{Blatter-et-al}

In summary, in this paper, we have extended the approach developed in
refs.  \cite{Ettouhami1,Ettouhami2} to the case of a flux liquid in the
presence of a random pinning potential. This approach, which is based on the use of
the conformation variables $\{{\bf r}_n(z)\}$ as the {\em true} dynamical
variables in terms of which a Gaussian approximation is taken, gives physically
more reasonable results \cite{Ettouhami1,Ettouhami2} than the boson mapping 
 \cite{Nelson-LeDoussal}
or other hydrodynamic approaches \cite{Radzihovsky-Frey} which, instead, use the
density as the basic dynamical variable of the vortex liquid. Within our approach,
we find that the only effect of the pinning potential in the static equilibrium case 
is to renormalize the tilt modulus and the confining potential of the internal modes of the flux 
lines, increasing their stiffness and reducing their thermal wandering.
In a similar fashion, we find that in presence of pinning, apart from the appearance of 
nonlinear KPZ terms and standard renormalization of the coefficients, the equation of motion of 
flux lines keeps the same form as in an unpinned vortex liquid. 
As a consequence, and unlike the hydrodynamic approximations,
 \cite{Nelson-Seung,Nelson-LeDoussal,Tauber-Nelson,Radzihovsky-Frey}
we find that the structure factor
$S({\bf r},z)$ has the same functional form as in a liquid of
interacting but unpinned flux lines, with suitably renormalized parameters.
Our formulation of the equilibrium dynamics of vortex liquids is in full agreement 
with the standard dynamical theory of classical liquids, and through
the pair distribution function $g_0(r)$ of the c.m. mode, takes into account
nontrivial correlations in the positions of flux lines. In
particular, we find that the long time dynamics of 
a liquid of interacting flux lines
is qualitatively similar to the dynamics of an ordinary classical liquid of 
hard rods with a renormalized diffusion coefficient $D$
(which is reduced by the interactions with respect to the free value $D_0$),
the only effect of internal fluctuations of flux lines at long time scales
being to reduce the structure factor of the vortex liquid through 
the introduction of a Debye-Waller-like thermal smearing term.

\acknowledgments

The author acknowledges stimulating interaction with Prof. Leo Radzihovsky,
Prof. Cristina Marchetti, and Dr. George Crabtree.
Most of this work was done while the author was still at the University 
of Colorado at Boulder, where he was supported in part by the
David and Lucile Packard Foundation. The final stages were done in 
Gainesville, and the author wishes to thank the Department of Physics, University of 
Florida, for financial support.

\appendix

\section{Effective Hamiltonian of the internal modes of flux lines}
\label{App:Heff}

In this Appendix, we show details of how we perform the average $\langle H^{(1)}_{dis}\rangle_0$, 
where $H^{(1)}_{dis}$ is the disorder part of the Hamiltonian of the internal modes of the flux lines, 
and where the average is taken with statistical weight $\exp(-\bar{H}^{(0)}/T)/\mbox{Tr}(\exp{-\bar{H}^{(0)}/T})$.
As can be seen from equation (\ref{Hexpansion}), the first term of the Hamiltonian $H^{(1)}_{dis}$ 
does not depend on the c.m. coordinates. Assuming long wavelength elastic distorsions of the flux lines, we shall  
write:  \cite{Landau-Elasticity}
\begin{eqnarray}
{\bf u}_{i}^a(z)-{\bf u}_{i}^a(z') \simeq (z-z')\partial_z{\bf u}_i(z)
\end{eqnarray}
upon which one can see that the first term on the 
right hand side of equation (\ref{Hexpansion}) can be written in the form
\begin{eqnarray}
\sum_{a=1}^p\sum_{i=1}^N\int dz \;\frac{1}{2}\,\delta{K}\;\big(\partial_z{\bf
u}_i^a(z)\big)^2
\end{eqnarray}  
where the long-wavelength disorder contribution $\delta{K}$ to the tilt modulus of the flux lines is given by
\begin{eqnarray}
\delta{K} = -\frac{1}{d_\perp T}\int_{-\infty}^\infty dz\; z^2\nabla_\perp^2\Delta({\bf r},z)\Big|_{{\bf r}=0}
\end{eqnarray}
and vanishes for an ultralocal (in $z$) disorder, with a variance
$\Delta({\bf r},z)\propto\delta(z)$.

Now, for the second and third terms of $H^{(1)}_{dis}$,
we need to evaluate averages of the form
\begin{eqnarray}
\langle f_{ij}^{ab}\rangle_0 = \big\langle f\big({\bf r}_{0i}^a - {\bf r}_{0j}^b\big)\big\rangle_0
\end{eqnarray}
where $f({\bf r})$ is an arbitrary function of the space variable ${\bf r}$. We have
\begin{eqnarray}
\langle f_{ij}^{ab}\rangle_0 & = & \bar{Z}^{-1}\int \prod_{c,k}d{\bf r}_{oi}^c\;f({\bf r}_{0i}^a - {\bf r}_{0j}^b)
\,\mbox{e}^{-\beta \bar{H}} 
\end{eqnarray}
where $\bar{Z}=\mbox{Tr}\big(\exp(-\beta\bar{H})\big)$. Separating out the integrations over 
${\bf r}_{0i}^a$ and ${\bf r}_{0j}^b$, we obtain
\begin{eqnarray}
\langle f_{ij}^{ab}\rangle_0 & = & \int d{\bf r}_{0i}^a\,d{\bf r}_{0j}^b\;
f({\bf r}_{0i}^a-{\bf r}_{0j}^b)\times
\nonumber\\
&\times& \bar{Z}^{-1}\int \prod_{c,k}{}^{'}d{\bf r}_{0k}^c\;\mbox{e}^{-\beta\bar{H}}
\label{avg1}
\end{eqnarray}
where the prime in on the product indicates that the variables ${\bf r}_{0i}^a$ and ${\bf r}_{0j}^b$
do not appear in the integration measure.
Now, two cases have to be distinguished~:

{\em (i)} If $a=b$, {\em i.e.} ${\bf r}_{0i}^a$ and ${\bf r}_{0j}^b$ belong to the same replica. Then 
the quantity in the second line on the right hand side of equation (\ref{avg1}) is given by
(see also ref.  \cite{Ettouhami1})
\begin{eqnarray}
\frac{\rho^2}{N(N-1)}\;g_0({\bf r}_{0i}^a - {\bf r}_{0j}^a)
\end{eqnarray}
where $g_0({\bf r})$ is the pair distribution function of the c.m. of a given
replica in the flux liquid.
This leads to the following result for $\langle f_{ij}^{aa}\rangle_0$
\begin{eqnarray}
\langle f_{ij}^{aa}\rangle_0 = \frac{\rho}{N-1}\int d{\bf r}\;f({\bf r})g_0({\bf r})
\label{faa}
\end{eqnarray}

{\em (ii)} If $a\neq b$ (${\bf r}_{0i}^a$ and ${\bf r}_{0j}^b$ belong to different replicas),
then the quantity on the second line of equation (\ref{avg1}) is now given by
\begin{eqnarray}
\frac{\rho^2}{N^2}\;\tilde{g}_0({\bf r}_{0i}^a-{\bf r}_{0j}^b)
\end{eqnarray}
where $\tilde{g}_0({\bf r})$ is the pair distribution function of particles belonging to different 
replicas, equation (\ref{g0_aneqb}). This leads to the result
\begin{eqnarray}
\langle f_{ij}^{ab}\rangle_0 = \frac{\rho}{N}\int d{\bf r}\;f({\bf r})\tilde{g}_0({\bf r}) 
\label{fab}
\end{eqnarray}

Using the results (\ref{faa})-(\ref{fab}) to take the average of the second and third term on the 
right hand side of equation 
(\ref{Hexpansion}),  and rearraging the resulting sums, we obtain, after a few 
manipulations, the result (\ref{Heff_dis}) of the text.

\section{Hartree Approximation for the equilibrium dynamics of
undisordered vortex liquids}
\label{App:Hartree}

In this Appendix, we show how the result of Sec. \ref{secEquilibrium} for the
effective ``mass'' term $\mu$, where we used a simple Taylor expansion in the
flux-line displacements $\{{\bf u}\}$, can be generalized to take into
account the potentially large excursions of the vortices around their
c.m. positions which are possible in a vortex liquid. Here we shall use a
Hartree approximation, which is analogous to the approach introduced a long
time ago in the context of spin glasses  \cite{Sompolinsky}, and was extended
more recently to the spherical $p$-spin model  \cite{Crisanti} and to
fluctuating manifolds in random media  \cite{Kinzelbach}.
We begin by rewriting the interaction part ${\cal A}_{int}$ of
Eq. (\ref{Aint}) in the form
\begin{eqnarray}
{\cal A}_{int} = {\cal A}_{int}^{(0)} + {\cal A}_{int}^{(u)}
\end{eqnarray}
where
\begin{eqnarray}
{\cal A}_{int}^{(0)} & = & \sum_{i\neq j}\int dz\,dt\int_{\bf q} V({\bf q})iq_\alpha\,i\tilde{r}_{0i\alpha}(t)
\,\mbox{e}^{i{\bf q}\cdot[{\bf r}_{0i}(t)-{\bf r}_{0j}(t)]}
\nonumber\\
&\times&\mbox{e}^{i{\bf q}\cdot[{\bf u}_i(z,t)-{\bf u}_j(z,t)]}
\end{eqnarray}
is the c.m. part of the interacting action, and where
%\begin{widetext}
\begin{eqnarray}
{\cal A}_{int}^{(u)} &=& \sum_{i\neq j}\int\!\! dz\,dt\!\!\int_{\bf q} V({\bf q})iq_\alpha\,i\tilde{u}_{i\alpha}(z,t)
\mbox{e}^{i{\bf q}\cdot[{\bf r}_i(z,t)-{\bf r}_j(z,t)]},
\nonumber\\
& = & \sum_{i\neq j}\int dz\,dt\int_{\bf q} V({\bf q})iq_\alpha\,i\tilde{u}_{i\alpha}(z,t)
\mbox{e}^{i{\bf q}\cdot[{\bf r}_{0i}(t)-{\bf r}_{0j}(t)]}
\nonumber\\
&\times&\mbox{e}^{i{\bf q}\cdot[{\bf u}_i(z,t)-{\bf u}_j(z,t)]}
\end{eqnarray}
%\end{widetext}
is the internal modes contribution to ${\cal A}_{int}$. In the c.m. piece, we get rid of the $\{{\bf u}\}$ dependence 
by making the replacement
\begin{eqnarray}
\mbox{e}^{i{\bf q}\cdot[{\bf u}_i(z,t)-{\bf u}_j(z,t)]} \rightarrow
\mbox{e}^{-\frac{1}{2}\,q_\alpha q_\beta \phi_{ij}^{\alpha\beta}(z,t;z,t)}
\end{eqnarray}
where we defined the correlation function 
${\phi}_{i,j}^{\alpha\beta}(z,t;z',t')=\langle[u_{i\alpha}(z,t)-u_{j\alpha}(z',t')]
[u_{i\beta}(z,t)-u_{j\beta}(z',t')]\rangle$. This yields, for the c.m. part ${\cal A}_{int}^{(0)}$~:
\begin{equation}
{\cal A}_{int}^{(0)} \simeq \sum_{i\neq j}\int{dt}\int_{\bf q} 
\tilde{V}({\bf q})iq_\alpha\,iL\tilde{r}_{0i\alpha}(t)
\;\mbox{e}^{i{\bf q}\cdot[{\bf r}_{0i}(t)-{\bf r}_{0j}(t)]}
\end{equation}
with the effective interaction potential {\em per unit length} between vortices
\begin{eqnarray}
\tilde{V}({\bf q}) = V({\bf q})\,\mbox{e}^{-\frac{1}{2}\,q_\alpha q_\beta \phi_{ij}^{\alpha\beta}(z,t;z,t)}
\label{tildeV}
\end{eqnarray}
which is smeared with respect to the original potential $V(q)$ by th thermal fluctuations of the internal modes of flux 
lines.

We now turn our attention to the relatively more involved task of constructing a self consistent Gaussian 
approximation for ${\cal A}_{int}^{(u)}$.
Following Scheidl (who derived a similar self consistent approach for a flux line lattice pinned by 
disorder  \cite{Scheidl}),
we expand the exponential with respect to the displacements and contract the fields in all possible ways pairwise
until one or two fields remain uncontracted. For even and odd terms in the displacement, this yields:
\begin{widetext}
\begin{eqnarray}
i\tilde r_{i\alpha}(z,t)&\times&\frac{1}{(2n)!}\{i{\bf q}\cdot[{\bf u}_i(z,t)-{\bf u}_j(z,t)]\}^{2n} \rightarrow
i{\tilde r}_{i^\alpha}(z,t)\frac{1}{n!}\{-\frac{1}{2}q_\alpha q_\beta{\phi}_{i,j}^{\alpha\beta}(z,t;z,t)\}^n
\nonumber\\
i\tilde r_{i^\alpha}(z,t)&\times&\frac{1}{(2n+1)!}\{i{\bf q}\cdot[{\bf u}_i(z,t)-{\bf u}_j(z,t)]\}^{2n+1}
\rightarrow 
i{\tilde r}_{i^\alpha}(z,t)\,\{i{\bf q}\cdot[{\bf u}_i(z,t)-{\bf u}_j(z,t)]\}\,
\frac{1}{n!}\{-\frac{1}{2}\,q_\alpha q_\beta{\phi}_{i,j}^{\alpha\beta}(z,t;z,t)\}^n
\nonumber
\end{eqnarray}
Resummation yields:
\begin{eqnarray}
{\cal A}_{int}^{(u)} \simeq \sum_{i\neq j}\int dz\,dt \;i{\tilde r}_i^\alpha(z,t)\,\big\{V^\alpha_{i,j}(z,t;z,t)+
[u_{i\beta}(z,t)-u_{j\beta}(z,t)]\,V^{\alpha,\beta}_{i,j}(z,t;z,t)\big\}
\end{eqnarray}
where we defined for convenience:
\begin{subequations}
\begin{eqnarray}
V^\alpha_{i,j}(z,t;z,t) & = & \int_{\bf q} V({\bf q})iq_\alpha
\,\mbox{e}^{i{\bf q}\cdot[{\bf r}_{0i}(t)-{\bf r}_{0j}(t)]}
\mbox{e}^{-\frac{1}{2}q_{\alpha}q_{\beta}{\phi}_{i,j}^{\alpha\beta}(z,t;z,t)}
\label{V1}\\
V^{\alpha,\beta}_{i,j}(z,t;z,t) & = & \int_{\bf q} V({\bf q})iq_{\alpha}iq_{\beta}
\mbox{e}^{i{\bf q}\cdot[{\bf r}_{0i}(t)-{\bf r}_{0j}(t)]}\,
\mbox{e}^{-\frac{1}{2}q_{\alpha}q_{\beta}{\phi}_{i,j}^{\alpha\beta}(z,t;z,t)}
\label{V2}
\end{eqnarray}
\end{subequations}
\end{widetext}\noindent
The term $V_{i,j}^{\alpha,\beta}(t)$ represents a force acting on particle $i$ and arising from particle $j$,
which is proportional to the small displacement $u_j^\alpha$. 

With hindsight from the results of Section \ref{secEquilibrium}, we shall
assume that 
$\langle u_{i\alpha}(z,t)u_{j\beta}(z,t)\rangle
=\delta_{ij}\delta_{\alpha\beta}\,\langle u_{i\alpha}^2\rangle$, which 
implies that, for $i\neq j$,
\begin{eqnarray}
{\phi}_{i,j}^{\alpha\beta}(z,t;z,t) = \frac{2}{d_\perp}\delta_{\alpha\beta}\langle u^2\rangle 
\end{eqnarray}
The effective smeared interaction potential of Eq. (\ref{tildeV}) becomes
\begin{eqnarray}
\tilde{V}({\bf q}) = V({\bf q})\,\mbox{e}^{-\frac{1}{d_\perp}\langle u^2\rangle}
\end{eqnarray}
The quantities $V_{ij}^\alpha$ and $V_{ij}^{\alpha\beta}$ in Eqs. (\ref{V1})-(\ref{V2}), on the other hand, are now 
given by~:
\begin{subequations}
\begin{eqnarray}
V^\alpha_{i,j}(z,t;z,t) & = & \int_{\bf q} V({\bf q})iq_\alpha
\,\mbox{e}^{i{\bf q}\cdot[{\bf r}_{0i}(t)-{\bf r}_{0j}(t)]-\frac{q^2}{2d_\perp}\langle u^2\rangle}
\nonumber
\\
V^{\alpha,\beta}_{i,j}(z,t;z,t) & = & \int_{\bf q} V({\bf q})iq_{\alpha}iq_{\beta}
\mbox{e}^{i{\bf q}\cdot[{\bf r}_{0i}(t)-{\bf r}_{0j}(t)]-\frac{q^2}{2d_\perp}\langle u^2\rangle}
\nonumber
\end{eqnarray}
\end{subequations}
Taking the average of ${\cal A}_{int}$ over the center of mass positions with statistical weight $\mbox{e}^{-A_0}$, we 
obtain the following effective interaction action:
\begin{eqnarray}
{\cal A}_{int}^{(u)} &\simeq& \sum_{i\neq j}\int dz\,dt \;i{\tilde u}_i^\alpha(z,t)\,\big\{ 
\tilde{V}^\alpha_{i,j}(z,t;z,t)+
\nonumber\\
& + & [u_{i\beta}(z,t)-u_{j\beta}(z,t)]\,\tilde{V}^{\alpha,\beta}_{i,j}(z,t;z,t)\big\}
\label{AuHartree}
\end{eqnarray}
where now
\begin{subequations}
\begin{eqnarray}
V^\alpha_{i,j}(z,t;z,t) & = & \int_{\bf q} V({\bf q})iq_\alpha
\;g_0({\bf q})\,
\mbox{e}^{-\frac{q^2}{2d_\perp}\langle u^2\rangle}
\nonumber\\
\label{newV1}
\\
V^{\alpha,\beta}_{i,j}(z,t;z,t) & = & \int_{\bf q} V({\bf q})iq_{\alpha}iq_{\beta}
\;g_0({\bf q})\,
\mbox{e}^{-\frac{q^2}{2d_\perp}\langle u^2\rangle}
\nonumber\\
\label{newV2}
\end{eqnarray}
\end{subequations}
where we used the fact that
\begin{eqnarray}
\Big\langle \mbox{e}^{i{\bf q}\cdot[{\bf r}_{0i}(t)-{\bf r}_{0j}(t)]} \Big\rangle_0 =
\frac{\rho}{(N-1)}\;g_0({\bf q})
\end{eqnarray}
It is not difficult to see that $\tilde{V}^\alpha$ vanishes due to spherical symmetry of the interaction potential,
$V({\bf q})=V(q)$. Putting together all terms in Eq. (\ref{AuHartree}), one obtains that the effective action for the 
internal modes of flux lines can be again rewritten in the quadratic form of Eqs. (\ref{Aeff})-(\ref{kernelF}), 
with now the mass coefficient $\mu$ given by the self consistent equation
\begin{eqnarray}
\mu = -\frac{\rho}{d_\perp}\int_{\bf q} q^2V({\bf q})g_0({\bf q})
\mbox{e}^{-q^2 \langle u^2\rangle/d_\perp}
\label{defmueff}
\end{eqnarray}
which is identical to the result obtained in ref.  \cite{Ettouhami2} 
using a static variational approach.

%%%%%%%%%%%%%%%%%

\section{Gaussian hydrodynamics of flux lines revisited}
\label{App:Hydrodynamics}

In this Section, we revisit the Gaussian hydronamic formulation of the statistical 
mechanics of flux liquids, which will make it easier for us to compare the results 
of our microscopic approach to the results of previous publications 
\cite{Nelson-Seung,Nelson-LeDoussal,Tauber-Nelson,Benetatos-Marchetti},
which were mostly based
on macroscopic coarse-graining methods. We shall first consider the case of a liquid 
of rigid flux lines, before considering the general case of a liquid of flexible vortices.

\subsection{Hydrodynamics of rigid flux lines}
\label{Sub:rigid_lines}

Let us consider a system of rigid flux lines, described by the following Hamiltonian:
\begin{equation}
H = \frac{1}{2}\sum_{i,j} V_0({\bf r}_i-{\bf r}_j) + \sum_{i} V_{ext}({\bf r}_i) \,,
\end{equation}
where $V_0({\bf r})=LV({\bf r})$, and where we introduced a one-body external 
potential $V_{ext}({\bf r})$. The hydrodynamics of the liquid described by the 
above Hamiltonian is constructed in a standard way  \cite{Chaikin-Lubensky} as follows. First, one
introduces a variational Hamiltonian, consisting of the sum of non-interacting single-vortex Hamiltonians:
\begin{equation}
H_v = \sum_{i=1}^N H_1({\bf r}_i)\,,
\label{Hv}
\end{equation}
with the Hamiltonian $H_1$ to be determined by minimization of the variational free energy $F_\rho$
(the significance of the subscript $``\rho"$ will become clear shortly), which is given by:
\begin{equation}
F_\rho = -T\ln Z_v + \langle H - H_v \rangle_v \,.
\label{defFrho}
\end{equation} 
In the above expression, $\langle \cdots \rangle_v$ denotes averaging with statistical weight
$e^{-\beta H_v}/Z_v$, and $Z_v$ is the
partition function associated with the Hamiltonian $H_v$, and is given by:
\begin{eqnarray}
Z_v & = & \int d{\bf r}_1\cdots d{\bf r}_N\; e^{-\beta\sum_{i=1}^N H_1({\bf r}_i)} \,,
\nonumber\\
& = & \Big(\int d{\bf r} \; e^{-\beta H_1({\bf r})}\Big)^N \,.
\label{Zv}
\end{eqnarray}
In the same way, it is easy to show that:
\begin{eqnarray}
\langle H_v \rangle_v = \frac{N\int d{\bf r}\; H_1({\bf r})\,e^{-\beta H_1({\bf r})}}
{\int d{\bf r}\; e^{-\beta H_1({\bf r})}} \,.
\label{Hv_avg}
\end{eqnarray}
Now, the variational average of the density operator 
$\rho({\bf r})=\langle \sum_{i=1}^N\delta({\bf r}-{\bf r}_i)\rangle_v$ 
is given by:
\begin{equation}
\rho({\bf r}) = \frac{N e^{-\beta H_1({\bf r})}}{\int d{\bf r}\; e^{-\beta H_1({\bf r})}}\,,
\label{rho_v}
\end{equation}
so that:
\begin{equation}
\ln\Big(\frac{\rho({\bf r})}{N}\Big) = -\beta H_1({\bf r}) - \ln\Big(\int d{\bf r}\;e^{-\beta H_1({\bf r})}\Big) \,.
\end{equation}
Hence:
\begin{eqnarray}
\int d{\bf r}\; T\rho({\bf r})\ln\Big(\frac{\rho({\bf r})}{N}\Big) & = & 
- \frac{N\int d{\bf r}\; H_1({\bf r})\,e^{-\beta H_1({\bf r})}}{\int d{\bf r}\; e^{-\beta H_1({\bf r})}} 
\nonumber\\
& - & NT \ln\Big(\int d{\bf r}\;e^{-\beta H_1({\bf r})}\Big) \,.
\nonumber\\
\end{eqnarray}
Comparing the above equation with Eqs. (\ref{Zv}) and (\ref{Hv_avg}), we see that:
\begin{eqnarray}
\int d{\bf r} \; T\rho({\bf r})\ln\Big(\frac{\rho({\bf r})}{N}\Big) = -T\ln Z_v - \langle H_v\rangle_v\,. 
\label{entropicterm}
\end{eqnarray}
On the other hand, it is not difficult to show that the variational average of the original Hamiltonian $H$
is given by:
\begin{eqnarray}
\langle H\rangle_v & = & \frac{1}{2}\int d{\bf r}d{\bf r}'\; \rho({\bf r})V_0({\bf r}-{\bf r}')\rho({\bf r}')
\nonumber\\
& + & \int d{\bf r}\; V_{ext}({\bf r})\rho({\bf r})\,.
\label{interactionterm}
\end{eqnarray}
Collecting together the results (\ref{entropicterm}) and (\ref{interactionterm}), 
it follows that the variational free energy of Eq. (\ref{defFrho}) can be
written as a functional of the averaged density $\rho({\bf r})$ 
(hence the subscript $``\rho"$ in $F_{\rho}$), and is given by:
\begin{eqnarray}
F_\rho & = & \int d{\bf r} \; T\rho({\bf r})\ln\Big(\frac{\rho({\bf r})}{N}\Big) +
\int d{\bf r}\; V_{ext}({\bf r})\rho({\bf r})
\nonumber\\
& + & \frac{1}{2}\int d{\bf r}d{\bf r}'\; \rho({\bf r})V_0({\bf r}-{\bf r}')\rho({\bf r}') \,.
\label{F_rho_rigid}
\end{eqnarray}
The first term on the right hand side of the above equation is the entropic contribution to the free
energy of the liquid, which is qualitatively important (we shall see below that this term modifies the 
temperature dependence of the density response and correlation functions --- 
see Eqs. (\ref{Eq:result_chi_rigid}) and (\ref{Snn})), but which
has been systematically overlooked in previous studies of flux-line liquids.
 \cite{Nelson-Seung,Nelson-LeDoussal,Tauber-Nelson,Benetatos-Marchetti}
We now need to minimize the variational free energy $F_\rho$ with respect to the density $\rho({\bf r})$,
which may be thought of as an ``order parameter", 
with the constraint that the total number of particles $N$ is held fixed. This amounts to minimizing 
the variational version of the grand potential: 
\begin{equation}
\Omega_\rho = F_\rho-\mu\int d{\bf r}\,\rho({\bf r}) \,,
\label{Eq:Omega}
\end{equation} 
with respect to variations in the density $\rho({\bf r})$, using  \cite{Evans}:
\begin{equation}
\frac{\delta\Omega_\rho}{\delta\rho({\bf r})} = 0 \,,
\end{equation} 
with the Lagrange multiplier $\mu$ (chemical potential) in Eq. (\ref{Eq:Omega}) fixed by the condition 
$\int d{\bf r}\,\rho({\bf r})=N$. This minimization procedure
leads  \cite{Chaikin-Lubensky} to an expression for the one-body density $\rho({\bf r})$ 
which is identical to Eq. (\ref{rho_v}) above, with $H_1$ given by:
\begin{equation}
H_1({\bf r}) = V_{ext}({\bf r}) + \int d{\bf r}' \,V_0({\bf r}-{\bf r}')\rho({\bf r}')\,.
\label{resH1}
\end{equation}
Now, the density response function is given by:  \cite{Chaikin-Lubensky}
\begin{equation}
\chi({\bf r},{\bf r}') = - \frac{\delta\rho({\bf r})}{\delta V_{ext}({\bf r}')} \,.
\label{Eq:def_chi_rigid}
\end{equation}
Taking the functional derivative of $\rho({\bf r})$ in Eq. (\ref{rho_v}) with respect to $V_{ext}({\bf r}')$, 
one can easily show that:
\begin{eqnarray}
\chi({\bf r},{\bf r}') & = & \beta\rho({\bf r})\frac{\delta H_{1}({\bf r})}{\delta V_{ext}({\bf r}')}
\nonumber\\
& - & \frac{\beta\rho({\bf r})}{N} \int d{\bf r}_1 \;\rho({\bf r}_1)\frac{\delta H_1({\bf r}_1)}{\delta V_{ext}({\bf r}')}\,,
\nonumber\\
&\simeq& \beta\rho({\bf r})\frac{\delta H_1({\bf r})}{\delta V_{ext}({\bf r}')} \,,
\label{Eq:int_chi}
\end{eqnarray}
where, in going from the first to the second equality, we discarded a term proportional to 
$1/N$ which vanishes in the thermodynamic $N\to \infty$ limit. Now, using the result
(\ref{resH1}) for $H_1$ into Eq. (\ref{Eq:int_chi}), we obtain:
\begin{equation}
\chi({\bf r},{\bf r}') = \beta\rho({\bf r})\big[\delta({\bf r}-{\bf r}') - 
\int d{\bf r}'' \; V_0({\bf r}-{\bf r}'')\chi({\bf r}'',{\bf r}')\big]\,.
\label{chi1}
\end{equation}
In the absence of an external one-body potential (i.e. when $V_{ext}=0$), 
the system is translationally invariant, and
$\chi({\bf r},{\bf r}')=\chi({\bf r}-{\bf r}')$. Taking the Fourier transform of Eq. (\ref{chi1}) 
then leads to the result:
\begin{equation}
\chi({\bf q}) = \frac{1}{V_0({\bf q}) + T/\rho}\,.
\label{Eq:result_chi_rigid}
\end{equation}
The $T/\rho$ term in the denominator originates from the $\rho\ln\rho$ entropic term 
in expression (\ref{F_rho_rigid}) of the free energy. As we mentioned above, this term, 
which determines the temperature variation of the density response function, was totally 
ignored in previous studies of flux liquids. 
 \cite{Nelson-Seung,Nelson-LeDoussal,Tauber-Nelson,Benetatos-Marchetti}
From the above response function, the Ursell function 
$S_{nn}({\bf r})=\langle\rho({\bf r})\rho({\bf 0})\rangle -\rho^2$
(we use the terminology and notation of ref.  \cite{Chaikin-Lubensky}) is given by:
\begin{equation}
S_{nn}({\bf r}) = T\chi({\bf r})\,,
\end{equation}
which gives, in Fourier space:
\begin{equation}
S_{nn}({\bf q}) = T\chi({\bf q}) = \frac{T}{V_0({\bf q}) + T/\rho}\,.
\label{Snn}
\end{equation}
The above Ursell function can be obtained directly from Eq. (\ref{F_rho_rigid})
by writing the density $\rho({\bf r})$ as $\rho({\bf r})=\rho + \delta\rho({\bf r})$,
and expanding $F_\rho$ up to quadratic order in the density fluctuation $\delta\rho({\bf r})$.
Taking the Gaussian average of the product $\delta\rho({\bf r})\delta\rho({\bf 0})$
by integrating directly over the density fluctuation field $\delta\rho({\bf r})$,
\begin{equation}
\langle \delta\rho({\bf r},z)\delta\rho({\bf 0},0) \rangle =
\frac{\int [d(\delta\rho({\bf r},z))]\,\delta\rho({\bf r},z)\delta\rho({\bf 0},0)\;e^{-\beta F_\rho}}
{\int [d(\delta\rho({\bf r},z))]\;e^{-\beta F_\rho}}\,,
\label{StatAvgF_rho}
\end{equation}
leads directly to the result (\ref{Snn}). Note however that this last procedure to obtain the Ursell function 
is an {\em ad hoc} one (that is valid only because $\Omega_\rho$ is stationary with respect to variations
of the density $\rho({\bf r})$ around its equilibrium value $\rho$), 
and that the most systematic and justifiable way to obtain $S_{nn}$ in this
variational formulation of hydrodynamics is through extremizing the grand potential $\Omega_\rho$ 
with respect to the density (a step that is rigourosouly
exact \cite{Evans}), and then using the resulting variational Hamiltonian to find the density response function and 
hence $S_{nn}({\bf q})$, as we did in Eqs. (\ref{F_rho_rigid})-(\ref{Snn}).
We insist that in the latter method no integration over density variables (as in Eq. (\ref{StatAvgF_rho})) 
is performed, and we thus see that Eq. (\ref{StatAvgF_rho}) is by no means a necessary step
to obtain the Ursell function.

% rationale behind this method is that

\subsection{Hydrodynamics of flexible flux lines}

We now want to generalize the approach of the above Subsection 
to flexible flux lines. For the sake of homogeneity with 
the rest of the paper, we shall use the same Hamiltonian as in the text, namely:
\begin{eqnarray}
H & = & \sum_{i=1}^N\int dz \;\big[\frac{1}{2}\,K\Big(\frac{d{\bf r}_i}{dz}\Big)^2 
+ V_{ext}({\bf r}_i(z),z)\big]
\nonumber\\
&+& \frac{1}{2}\sum_{i,j}\int dz\; V({\bf r}_i(z)-{\bf r}_j(z))\,,
\end{eqnarray}
where we again, by analogy with the previous Subsection,
have introduced a one-body external potential $V_{ext}({\bf r},z)$.
In an obvious generalization of Eq. (\ref{Hv}) to continuous systems, we shall use the following variational
Hamiltonian:
\begin{equation}
H_v = \sum_{i=1}^N H_1[{\bf r}_i(z)] \,,
\label{Eq:H1}
\end{equation}
where $H_1$ now is a functional of the entire trajectory ${\bf r}_i(z)$ of the $i$-th flux line.
The corresponding partition function is given by:
\begin{eqnarray}
Z_v & = & \int [d{\bf r}_1(z)]\cdots[d{\bf r}_N(z)] \;
\exp\Big(-\beta \sum_{i=1}^N H_1[{\bf r}_i(z)]\Big)\,,
\nonumber\\
& = & \Big(\int [d{\bf r}(z)]\;e^{-\beta H_1[{\bf r}(z)]}\Big)^N
\end{eqnarray}
The variational average of $H_v$ on the other hand is given by:
\begin{equation}
\langle H_v\rangle_v = \frac{N\int [d{\bf r}(z)]\;H_1[{\bf r}(z)]\,e^{-\beta H_1[{\bf r}(z)]}}
{\int [d{\bf r}(z)]\;e^{-\beta H_1[{\bf r}(z)]}} \,,
\end{equation}
in total analogy with Eq. (\ref{Hv_avg}).
Let us now define the operator:
\begin{equation}
\hat\rho[{\bf r}(z)] = \sum_{i=1}^N \prod_z \delta\big({\bf r}(z)-{\bf r}_i(z)\big) \,,
\label{Eq:trajectory-density}
\end{equation}
whose statistical average $\langle\hat\rho[{\bf r}(z)]\rangle$ may be interpreted as the average probability
density for finding a vortex with a specific trajectory ${\bf r}(z)$ in the system. Now, the variational
average $\rho[{\bf r}(z)]=\langle\hat\rho[{\bf r}(z)]\rangle_v$ of this trajectory-density operator 
is given by:
\begin{equation}
\rho[{\bf r}(z)] = \frac{N\,e^{-\beta H_1[{\bf r}(z)]}}{\int [d{\bf r}(z)]\;e^{-\beta H_1[{\bf r}(z)]}}\,.
\label{Eq:rho-trajectory}
\end{equation}
Forming $T\rho[{\bf r}(z)]\ln(\rho[{\bf r}(z)]/N)$, and then taking the functional integral over the trajectory ${\bf r}(z)$,
one can again easily show that:
\begin{equation}
\int [d{\bf r}(z)]\;T\rho[{\bf r}(z)]\ln\Big(\frac{\rho[{\bf r}(z)]}{N}\Big) = -T\ln Z_v - \langle H_v \rangle_v. 
\end{equation}
Now, the variational average of $H$ can be expressed in terms of the trajectory density $\rho[{\bf r}(z)]$,
and is given by:
\begin{eqnarray}
\langle H\rangle_v & = & \int dz \int [d{\bf r}(z)]\Big\{\Big[\frac{1}{2}\,K\Big(\frac{d{\bf r}}{dz}\Big)^2 
\nonumber\\
& + & V_{ext}({\bf r}(z),z)\Big]\rho[{\bf r}(z)]
\nonumber\\
& + & \rho[{\bf r}(z)]V\big({\bf r}(z) - {\bf r}'(z)\big)\rho[{\bf r}'(z)]\Big\}\,,
\end{eqnarray}
and hence we obtain that the variational free energy 
$F_\rho = -T\ln Z_v + \langle H-H_v\rangle_v$ can be written in the form:
\begin{eqnarray}
F_\rho & = & \int dz \Big\{\int [d{\bf r}(z)]\;T\rho[{\bf r}(z)]\ln\Big(\frac{\rho[{\bf r}(z)]}{N}\Big)
\nonumber\\
& + &\int [d{\bf r}(z)] \Big[\frac{1}{2}\,K\Big(\frac{d{\bf r}}{dz}\Big)^2 
+ V_{ext}({\bf r}(z))\Big]\rho[{\bf r}(z)]
\nonumber\\
& + & \int\![d{\bf r}(z)][d{\bf r}'(z)] \rho[{\bf r}(z)]V\big({\bf r}(z) - {\bf r}'(z)\big)\rho[{\bf r}'(z)]
\Big\},
\nonumber\\
\end{eqnarray}
and is a generalized functional of the averaged trajectory density $\rho[{\bf r}(z)]$.
In the subsequent steps of the hydrodynamic method, which involve minimizing $F_\rho$ with respect to
the density $\rho[{\bf r}(z)]$, it is convenient to discretize the $z$ axis, with unit step $d$, and think of 
$\rho[{\bf r}(z)]$ as a function $\rho\big({\bf r}(0),{\bf r}(d),\ldots,{\bf r}(Md)\big)$ of the positions
$\big({\bf r}(0),{\bf r}(d),\ldots,{\bf r}(Md)\big)$ of $M+1$ particles interacting
with each other harmonically, with a spring constant $K/d$. 
The integration measure in this case can be defined as:
\begin{equation}
[d{\bf r}(z)] = \prod_{m=0}^M d{\bf r}(md) \,.
\end{equation}
Now, from Eq. (\ref{Eq:rho-trajectory}) we see that:
\begin{equation}
\int [d{\bf r}(z)]\; \rho[{\bf r}(z)] = N\,,
\label{Eq:constraint_rho(z)}
\end{equation}
and hence we see that the grand potential $\Omega_\rho=F_\rho-\mu N$
can be written in the form:
\begin{equation}
\Omega_\rho = F_\rho - \mu\int [d{\bf r}(z)]\; \rho[{\bf r}(z)] \,.
\end{equation}
In the same way as in the previous Subsection, extremization of $\Omega_\rho$ with respect to 
$\rho[{\bf r}(z)]=\rho\big({\bf r}(0),{\bf r}(d),\ldots,{\bf r}(Md)\big)$ 
(thought of as a ``generalized" order parameter) under the constraint (\ref{Eq:constraint_rho(z)})
leads to the following result for the variational Hamiltonian $H_1[{\bf r}(z)]$:
\begin{eqnarray}
H_1[{\bf r}(z)] & = & \int dz \Big\{\Big[\frac{1}{2}\,K\Big(\frac{d{\bf r}}{dz}\Big)^2 
+ V_{ext}({\bf r}(z))\Big]
\nonumber\\
& + & \int[d{\bf r}'(z)]\,V\big({\bf r}(z) - {\bf r}'(z)\big)\rho[{\bf r}'(z)]\Big\}.\quad
\label{Eq:resH1}
\end{eqnarray}
Like we did in the text, we now decompose the trajectory ${\bf r}(z)$ into c.m. and internal coordinates,
${\bf r}(z)={\bf r}_0 + {\bf u}(z)$, with ${\bf r}_0=\int_0^L dz\; {\bf r}(z)$ being the c.m. coordinate
of the trajectory ${\bf r}(z)$, and Taylor expand the interaction part in the small displacement ${\bf u}({\bf r})$:
\begin{eqnarray}
V\big({\bf r}(z) - {\bf r}'(z)\big) &=& V\big({\bf r}_0 - {\bf r}'(z)\big) 
\nonumber\\
&+& u_\alpha(z)\partial_{\alpha}V\big({\bf r}_0 - {\bf r}'(z)\big) 
\nonumber\\
&+& \frac{1}{2}u_\alpha(z)u_\beta(z)\partial_{\alpha}\partial_{\beta}V\big({\bf r}_0 - {\bf r}'(z)\big)\,.
\nonumber\\
\end{eqnarray}
The Hamiltonian $H_1$ becomes:
\begin{widetext}
\begin{eqnarray}
H_1[{\bf r}(z)] & = & \int dz \Big[\frac{1}{2}\,K\Big(\frac{d{\bf u}}{dz}\Big)^2 
+ V_{ext}({\bf r}(z))\Big]
+ \int dz \int [d{\bf r}'(z)] \; V\big({\bf r}_0 - {\bf r}'(z)\big)\rho[{\bf r}'(z)] +
\nonumber\\
& + & \int dz \,u_\alpha(z) \int[d{\bf r}'(z)]\,\partial_{\alpha}V\big({\bf r}_0 - {\bf r}'(z)\big)\rho[{\bf r}'(z)]
+ \frac{1}{2}\int dz\; u_\alpha(z)u_\beta(z) \int [d{\bf r}'(z)]
\,\partial_\alpha\partial_\beta V\big({\bf r}_0-{\bf r}'(z)\big) \rho[{\bf r}'(z)]\,.
\nonumber\\
\end{eqnarray}
\end{widetext}
Using the fact that $\int_0^L dz {\bf u}(z)=0$, we see that the third term on the right hand side of the above equation
vanishes, and therefore that $H_1[{\bf r}(z)]$ in the absence of an external potential ($V_{ext}=0$) can be written in the form:
\begin{equation}
H_1[{\bf r}(z)] = H_{1,c.m.}({\bf r}_0) + H_{1,u}[{\bf u}(z)]\,,
\label{H1decomp}
\end{equation}
where:
\begin{subequations}
\begin{eqnarray}
H_{1,c.m.}({\bf r}_0) & = & \int dz \int [d{\bf r}'(z)] V\big({\bf r}_0-{\bf r}'(z)\big)\,\rho[{\bf r}'(z)] \,,
\label{H1cm}
\nonumber\\
\\
H_{1,u}[{\bf u}(z)] & = & \int dz \,\Big[\frac{1}{2} K\Big(\frac{d{\bf u}}{dz}\Big)^2 + \frac{1}{2}\mu_{\alpha\beta} u_\alpha(z)u_\beta(z)\Big]\,.
\nonumber\\
\label{H1u}
\end{eqnarray}
\end{subequations}
In the above expression of $H_{1,u}[{\bf u}(z)]$, we defined the confining ``mass" tensor:
\begin{equation}
\mu_{\alpha\beta} = \int [d{\bf r}'(z)] \,\partial_\alpha\partial_\beta V\big({\bf r}_0-{\bf r}'(z)\big)\rho[{\bf r}'(z)] \,.
\label{mualphabeta}
\end{equation}
As defined above, the quantities $\mu_{\alpha\beta}$ depend on ${\bf r}_0$. 
We shall however verify {\em a posteriori} that in the homogeneous
liquid state this dependence drops out and the $\mu_{\alpha\beta}$'s reduce
to ordinary constants (and are in fact all equal to zero in the hydrodynamic limit).

An immediate consequence of the decomposition (\ref{H1decomp}) is that the density matrix also can be written in a decoupled form:
\begin{equation}
\rho[{\bf r}(z)] = \rho_{c.m.}({\bf r}_0) \,\rho_{u}[{\bf u}(z)] \,,
\label{rhodecomp}
\end{equation}
with:
\begin{eqnarray}
\rho_{c.m.}({\bf r}_0) & = & \frac{e^{-\beta H_{1,c.m.}({\bf r}_0)}}{\int d{\bf r}_0 \; e^{-\beta H_{1,c.m.}({\bf r}_0)}} \,,
\\
\rho_{u}[{\bf u}(z)] & = & \frac{e^{-\beta H_{1,u}[{\bf u}(z)]}}{\int [d{\bf u}(z)] \;e^{-\beta H_{1,u}[{\bf u}(z)]} } \,.
\end{eqnarray}
Now, if we replace the density matrix $\rho[{\bf r}(z)]$ by the decoupled form (\ref{rhodecomp}) back into expression (\ref{H1cm})
of $H_{1,c.m.}$, we obtain:
\begin{equation}
H_{1,c.m.}({\bf r}_0) = \int d{\bf r}_0' \;L\tilde{V}({\bf r}_0 -{\bf r}_0')\rho_{c.m.}({\bf r}_0')\,,
\label{Eq:res_H1cm}
\end{equation}
where we defined:
\begin{subequations}
\begin{eqnarray}
\tilde{V}({\bf r}) & = & \int[d{\bf u}'(z)]\; V\big({\bf r}-{\bf u}'(z)\big)\,\rho_u[{\bf u}'(z)]\,,
\label{Eq:tildeV}
\\
& = & \int_{\bf k} e^{i{\bf k}\cdot{\bf r} - \frac{1}{2}k^2\langle u^2\rangle}V({\bf k})\,,
\label{Eq:tildeV2}
\end{eqnarray}
\end{subequations}
and where, in going from the first to the second line of the last equation, we assumed that
the density matrix $\rho_u[{\bf u}'(z)]$ represents an isotropic Gaussian distribution
for the displacement field $\{{\bf u}'(z)\}$.
In Eq. (\ref{Eq:tildeV}), the integration measure $[d{\bf u}(z)]$ stands for:
\begin{equation}
[d{\bf u}(z)] = \prod_{m=0}^{M-1} d{\bf u}(md) \,.
\end{equation}
Note that the transformation from the variables $\{{\bf r}(md)\}$ ($m=0,\ldots,M$) to
the variables $\{{\bf r}_0,{\bf u}(md)\}$ ($m=0,\ldots,M-1$) being linear, it has a constant Jacobian, that
we shall henceforth ignore for simplicity.

Now, if we insert the decoupled form (\ref{rhodecomp}) of $\rho[{\bf r}(z)]$ into expression
(\ref{mualphabeta}), we obtain:
\begin{eqnarray}
\mu_{\alpha\beta} & = & \int d{\bf r}_0' \;\rho_{c.m.}({\bf r}_0)
\nonumber\\
&\times&\int [d{\bf u}'(z)] \;\partial_\alpha\partial_\beta V\big({\bf r}_0-{\bf r}_0'-{\bf u}'(z)\big)\rho_u[{\bf u}'(z)] \,,
\nonumber\\
& = & \int d{\bf r}_0' \;\partial_\alpha\partial_\beta \tilde{V}\big({\bf r}_0-{\bf r}_0'\big)\rho_{c.m.}({\bf r}_0') \,.
\label{Eq:mualphabeta2}
\end{eqnarray}
In a uniform flux liquid, the averaged c.m. density reduces to a constant, $\rho_{c.m.}({\bf r}_0')=\rho$,
and the above integral vanishes. This is an direct consequence of the coarse-graining procedure, for if instead of
the {\em averaged} c.m. density $\rho_{c.m.}({\bf r}_0')$, we were still dealing with the c.m. density {\em operator}
$\hat{\rho}_{c.m.}({\bf r}_0)=\sum_{i=1}^N\delta({\bf r}_{0}-{\bf r}_{0i})$, then Eq. (\ref{Eq:mualphabeta2})
would give:
\begin{eqnarray}
\mu_{\alpha\beta} = \sum_{i=1}^N V({\bf r}_0-{\bf r}_{0i}) \,,
\end{eqnarray}
which is very similar to the undisordered version of Eq. (\ref{Eq:mui}), which then would yield finite
and isotropic ``mass'' coefficients $\mu_{\alpha\beta}=\mu\delta_{\alpha\beta}$.

Now, if we calculate the structure factor $S({\bf r},z)$ by directly taking the average of 
$\langle\hat\rho({\bf r},z)\hat\rho({\bf 0},0)\rangle$ as in Eq. (\ref{avg-density}) and using 
the decoupled density matrix of Eq. (\ref{rhodecomp}), one can easily show that the structure 
factor $S({\bf q},z)$ of the flux-line liquid has the form given in Eqs. (\ref{generalS}) 
and (\ref{ourS}) of the text, with $\mu=0$, namely:
\begin{equation}
S({\bf q},z) = \rho^2g_0({\bf q})\,e^{-q^2\langle u^2\rangle/d_\perp} + \rho\exp\Big(-\frac{Tq^2}{2K}|z|\Big)\,.  
\label{Eq:Sqz1}
\end{equation}
We have:
\begin{equation}
\rho^2g_0({\bf q}) = (2\pi)^2\rho^2\delta({\bf q}) + S_{0,nn}({\bf q}) - \rho\,,
\label{Eq:def_S0}
\end{equation}
where $S_{0,nn}$ is the Ursell function of the c.m. mode
($S_{0,nn}({\bf r})=\langle\rho_{c.m}({\bf r})\rho_{c.m.}({\bf 0})\rangle-\rho^2$), 
which can be easily derived
from the c.m. effective Hamiltonian (\ref{Eq:res_H1cm}), following the 
same steps as in the previous Subsection, with the result:
\begin{equation}
S_{0,nn}({\bf q}) = \frac{T}{L\tilde{V}({\bf q})+T/\rho}\,.
\label{Eq:res_S0nn}
\end{equation}
Using the result (\ref{Eq:def_S0}), Eq. (\ref{Eq:Sqz1}) can be rewritten in the form:
\begin{eqnarray}
S({\bf q},z) &=& (2\pi)^2\rho^2\delta({\bf q}) + [S_{0,nn}({\bf q})-\rho]\,e^{-q^2\langle u^2\rangle/d_\perp}
\nonumber\\
&+& \rho\exp\Big(-\frac{Tq^2}{2K}|z|\Big) \,.
\label{Eq:S-S0}
\end{eqnarray}
Since $\mu=0$ in the hydrodynamic limit, the mean squared displacement $\langle u^2\rangle\propto L$, and hence
$e^{-q^2\langle u^2\rangle/d_\perp}$ is exponentially small for practically all values of $q$ such that $0 < q\leq 1/a$.
Hence the above expression of the structure factor becomes:
\begin{equation}
S({\bf q},z) \simeq (2\pi)^2\rho^2\delta({\bf q}) 
+ \rho\exp\Big(-\frac{Tq^2}{2K}|z|\Big) \,,
\end{equation}
which implies that the Ursell function of the vortex liquid is given by:
\begin{equation}
S_{nn}({\bf q},z) = \rho\exp\Big(-\frac{Tq^2}{2K}|z|\Big) \,,
\label{Eq:res_Snn}
\end{equation}
and is identical, in the hydrodynamic limit, with the Ursell function of an ideal gas of non-interacting flux lines.
We thus see that, while the present formulation of Gaussian hydrodynamics yields a structure factor that is of the 
correct general functional form, Eq. (\ref{Eq:Sqz1})
(unlike previous formulations of refs.  \cite{Nelson-Seung,Nelson-LeDoussal,Tauber-Nelson,Benetatos-Marchetti}),
it fails to produce a non-zero value for the confining ``mass" term,
with the consequence that the ensuing strong fluctuations of flux lines completely smear out
the effect of interactions between flux lines (the second term on the right hand side of Eq. (\ref{Eq:S-S0})).

The decomposition given in Eq. (\ref{Eq:H1}), which approximates the $N$-body Hamiltonian 
of the system by a sum of $N$ noninteracting one-body Hamiltonians, is not the only possible 
choice for the variational Hamiltonian $H_v$ of the hydrodynamic method. One other 
(but not necessarily equivalent) possible choice consists, in the discretized scheme 
where the $z$ axis is cut into $M+1$ equidistant slices, in using the ansatz:
\begin{equation}
H_v = \sum_{i=1}^N\sum_{m=0}^M h\big({\bf r}(md),md\big) \,,
\label{HvNelson}
\end{equation}
with the Hamiltonian of a single vortex ``element" $h\big({\bf r}(md), md\big)$ 
at height $z=md$ to be determined variationally. The above ansatz amounts to
assuming for the Hamiltonian $H_1[{\bf r}(z)]$ of the preceding paragraphs the following form:
\begin{eqnarray}
H_1[{\bf r}(z)] = \sum_{m=0}^{M} h\big({\bf r}(md),md\big) \,.
\label{H1Nelson}
\end{eqnarray}
The variational partition function $Z_v$ is now given by:
\begin{eqnarray}
Z_v = \prod_{m=0}^M\Big(\int d{\bf r}\;e^{-\beta h({\bf r},md)}\Big)^{N}\,,
\end{eqnarray}
while the variational average $\langle H_v\rangle_v$ is given by:
\begin{eqnarray}
\langle H_v\rangle_v = \frac{N\sum_{m=0}^M\int d{\bf r} \;h({\bf r},md)e^{-\beta h({\bf r},md)} }{\int d{\bf r}\;e^{-\beta h({\bf r},md)}}\,.
\end{eqnarray}
If we define the density operator at height $z=md$, $\hat\rho({\bf r},md)$, by:
\begin{equation}
\hat\rho({\bf r},md) = \sum_{i=1}^N \delta({\bf r}-{\bf r}_i(md))\,,
\end{equation}
then it follows that the averaged density $\rho({\bf r},md)=\langle\hat\rho({\bf r},md)\rangle_v$
is given by:
\begin{equation}
\rho({\bf r},md) = \frac{Ne^{-\beta h({\bf r},md)}}{\int d{\bf r}\; e^{-\beta h({\bf r},md)}}\,.
\end{equation}
Thus, here again we can write:
\begin{widetext}
\begin{eqnarray}
\int d{\bf r}\; T\rho({\bf r},md)\ln\Big(\frac{\rho({\bf r},md)}{N}\Big) = 
-\frac{N\int d{\bf r}\; h({\bf r},md)e^{-\beta h({\bf r},md)} }{\int d{\bf r}\; e^{-\beta h({\bf r},md)}}
- TN\ln\Big(\int d{\bf r}\;e^{-\beta h({\bf r},md)} \Big) \,.
\end{eqnarray}
\end{widetext}
Summing over $m$, we obtain:
\begin{equation}
\sum_{m=0}^M \int d{\bf r}\; T\rho({\bf r},md)\ln\Big(\frac{\rho({\bf r},md)}{N}\Big) =
-T\ln Z_v - \langle H_v\rangle_v \,.
\end{equation}
Taking the variational average $\langle H\rangle_v$ of the original Hamiltonian $H$, 
we finally find that the variational free energy $F_\rho=-T\ln Z_v + \langle H-H_v\rangle_v$ 
can be written in the form (we now switch back to a continuum notation):
\begin{eqnarray}
F_\rho & = & \int d{\bf r}dz\;\Big[\frac{T}{d}\rho({\bf r},z)\ln\Big(\frac{\rho({\bf r},z)}{N}\Big) + 
\frac{1}{2}\,K_1\,{\bf t}^2({\bf r},z)\Big]
\nonumber\\
& + & \frac{1}{2} \int d{\bf r} \, d{\bf r}'\int dz\; \rho({\bf r},z)V({\bf r}-{\bf r}')\rho({\bf r}',z)
\nonumber\\
& + & \int d{\bf r}dz\;V_{ext}({\bf r},z)\rho({\bf r},z) \,,
\label{F_rho_flexible}
\end{eqnarray}
where $K_1=K/\rho$, and where we defined the ``tilt" field operator  \cite{Nelson-LeDoussal,Tauber-Nelson}:
\begin{equation}
\hat{\bf t}({\bf r},z) = \sum_{i=1}^N\frac{d{\bf r}_i}{dz}\delta\big({\bf r}-{\bf r}_i(z)\big) \,.
\end{equation}
The tilting field ${\bf t}({\bf r},z)$ and the density field $\rho({\bf r},z)$ are not independent, but are
related to each other by the continuity equation  \cite{Nelson-LeDoussal,Tauber-Nelson,Benetatos-Marchetti}:
\begin{equation}
\partial_z\rho({\bf r},z) + \nabla_\perp \cdot {\bf t}({\bf r},z) = 0 \,. 
\label{constraint_rho_t}
\end{equation}
Nelson and coworkers  \cite{Nelson-LeDoussal,Benetatos-Marchetti} obtain density and tilt correlation 
functions by expanding $F_\rho$ of Eq. (\ref{F_rho_flexible}) to quadratic order in 
$\delta\rho({\bf r},z)=\rho({\bf r},z)-\rho$
(omitting, for some unstated reason, the entropic $\rho\ln\rho$ term), 
and calculating statistical averages in the manner of Eq. (\ref{StatAvgF_rho}), 
with the constraint (\ref{constraint_rho_t}) enforced. For example,
for the Ursell function $S_{nn}({\bf r},z)=\langle \delta\rho({\bf r},z)\delta\rho({\bf 0},0) \rangle$, these authors write:
\begin{widetext}
\begin{equation}
S_{nn}({\bf r},z) \rangle =
\frac{\int [d(\delta\rho({\bf r},z))]\int[d{\bf t}({\bf r},z)]\;\delta(\partial_z\rho({\bf r},z) + \nabla_\perp \cdot {\bf t}({\bf r},z))
\;\delta\rho({\bf r},z)\delta\rho({\bf 0},0)\;e^{-\beta F_\rho}}
{\int [d(\delta\rho({\bf r},z))]\int[d{\bf t}({\bf r},z)]\;\delta(\partial_z\rho({\bf r},z) + \nabla_\perp \cdot {\bf t}({\bf r},z))e^{-\beta F_\rho}}\,.
\label{Snnflexible}
\end{equation}
\end{widetext}
with the result:
\begin{equation}
S_{nn}({\bf q},q_z) = \frac{T q_\perp^2}{\big[V({\bf q}) + T/({\rho}d)\big]q_\perp^2 + K_1q_z^2}\,,
\label{resSnnNelson}
\end{equation}
(As we mentioned already earlier, in previous treatments the $T/\rho d$ term in the denominator of the 
above expression, which comes from the entropic $\rho\ln\rho$ term in $F_\rho$, is missing.)
Unfortunately, the Gaussian integration in Eq. (\ref{Snnflexible}) is {\em not} justifiable from 
the point of view of a variational approach. In other words, correlation functions 
obtained by using Eq. (\ref{Snnflexible}) cannot be reproduced by a standard variational method, 
which here would consist in extremizing the free energy $F_\rho$ with respect to 
the fields $\delta\rho({\bf r},z)$ and ${\bf t}({\bf r},z)$, 
with steps similar to those of Eqs. (\ref{F_rho_rigid})-(\ref{Snn}) of the previous Subsection. 
Indeed, as we have seen in the end of Subsection 
\ref{Sub:rigid_lines} above, 
the rationale behind the Gaussian averaging in Eq. (\ref{Snnflexible}) is that the free energy is 
(presumably) a functional of $\delta\rho({\bf r},z)$ and ${\bf t}({\bf r},z)$
(with the constraint (\ref{constraint_rho_t}) enforced)
that is stationary at thermal equilibrium.

Technically, the standard way to implement the constraint (\ref{constraint_rho_t}) in a variational procedure is through
the introduction of a (functional) Lagrange multiplier $\lambda({\bf r},z)$, whereby one defines the following, 
modified grand potential (the last term in this equation is simply $-\mu N$):
\begin{eqnarray}
\tilde{\Omega}_\rho  & = &  F_\rho + 
\int d{\bf r}dz\;\lambda({\bf r},z)\Big(\partial_z\rho({\bf r},z) + \nabla_\perp \cdot {\bf t}({\bf r},z)\Big)
\nonumber\\
& - &\frac{\mu}{L}\int d{\bf r}dz\; \rho({\bf r},z) \,.
\end{eqnarray}
Extremizing $\tilde{\Omega}_\rho$ with respect to $\rho({\bf r},z)$ and ${\bf t}({\bf r},z)$ leads to the following coupled equations:
\begin{subequations}
\begin{eqnarray}
\partial_z \lambda({\bf r},z) & = & \frac{T}{d}\Big(\ln\Big(\frac{\rho({\bf r},z)}{N}\Big)+1\Big) + V_{ext}({\bf r},z) 
\nonumber\\
& + & \int d{\bf r}'\; V({\bf r}-{\bf r}')\rho({\bf r}',z) + \frac{\mu}{L} \,,
\label{lambda1}
\\
{\bm\nabla}_\perp\lambda({\bf r},z) & = & K_1 {\bf t}({\bf r},z) \,.
\label{lambda2}
\end{eqnarray}
\end{subequations}
Eq. (\ref{lambda2}) is a vector equation of standard form, which amounts to finding the 
``electric potential" $\lambda({\bf r},z)$ associated
with the planar ``electric field" ${\bf t}({\bf r},z)$,
and has a well-defined solution in $\lambda$ if and only if 
$\nabla_\perp\times{\bf t}=0$. Since ${\bf t}({\bf r},z)$ is 
a randomly fluctuating field that does {\em not} necessarily
satisfy this last condition, we arrive at the very important 
conclusion that it is {\em not} legitimate to extremize
the grand potential with respect to the pair of vector fields $\{\rho({\bf r},z),{\bf t}({\bf r},z)\}$, 
and hence that it is {\em not} legitimate to calculate statistical averages using the procedure 
examplified in Eq. (\ref{Snnflexible}).

We thus see that the previous formulations of Gaussian hydrodynamics 
which lead to expressions for the Ursell function
of the form given in Eq. (\ref{resSnnNelson}) correspond to a convoluted and {\em ad hoc} attempt,
without any rationale other than hand-waving symmetry considerations, to generalize the variational 
Gaussian hydrodynamics of point particles to continuous systems. 
In fact, even if we ignore this lack of rationale and 
%charitably 
accept the use of Eq. (\ref{Snnflexible}), 
one other source of inaccuracy of the previous formulations
of Gaussian hydrodynamics lies the underlying Hamiltonians of Eqs. (\ref{HvNelson})-(\ref{H1Nelson}),
which these theories are all implicitly based on. These Hamiltonians indeed represent 
a very crude approximation to the Hamiltonian of a single flux line in the first place, 
since $H_1[{\bf r}(z)]$ involves relatively strong (harmonic) interactions between flux line segments 
while Eq. (\ref{H1Nelson}) models a single flux line as a superposition of non-interacting elements. 
In fact, a necessary condition for the applicability of the Gaussian hydrodynamic approach is that 
the interactions be weak.  \cite{Chaikin-Lubensky}
For continuous systems with constituent parts interacting strongly, a correct formulation of 
Gaussian hydrodynamics must take these strong elastic interactions into account as exactly as possible, 
as we did in Eqs. (\ref{Eq:H1})-(\ref{Eq:resH1}), for otherwise one may obtain abnormal behaviour for 
$z$ correlations in the system (for example, the Ursell function $S_{nn}({\bf q},z)$ obtained from 
Eq. (\ref{resSnnNelson}) decays more rapidly than the corresponding quantity for an ideal gas of 
non-interacting flux lines, which is very surprizing, as discussed in detail in ref.  \cite{Ettouhami2}).

\subsection{Gaussian hydrodynamics of flexible flux-lines in presence of disorder}
\label{App:hydrodynamics_disorder}

We now generalize the formulation of Gaussian hydrodynamics that we developed in the previous Subsection
to flexible flux-lines in presence of disorder. Since the disordered case involves only minor technical 
modifications of the undisordered hydrodynamics, we shall only give the salient features of the calculation,
leaving out the (obvious) technical details.
Our starting point is the replicated Hamiltonian of Eq. (\ref{Hbar}),
which we rewrite here for clarity (we remind the reader that $p$ denotes the total number of replicas):
\begin{eqnarray} 
\bar{H}  & \!=\! & \sum_{a=1}^p\sum_{i=1}^N\int\!\! dz\,\frac{1}{2}\big\{
K\Big(\frac{d{\bf r}_i^a}{dz}\Big)^2 \!\! +
\sum_{j(\neq i)} V\big({\bf r}_i^a(z)\!-\!{\bf r}_j^a(z)\big)\big\}
\nonumber\\ 
&-& \frac{1}{2T}\sum_{a,b=1}^p\sum_{i,j=1}^N\int\!\!dz\,dz'
\,\Delta\big({\bf r}_i^a(z) \!-\! {\bf r}_j^b(z');z-z'\big)\,.
\nonumber\\
\label{Hbar-App}
\end{eqnarray}
Rewriting the above Hamiltonian in terms of the trajectory density operator of Eq. (\ref{Eq:trajectory-density}),
and using a variational ansatz for the total Hamiltonian of the system of the form:
\begin{equation}
H_v = \sum_{a=1}^p \sum_{i=1}^N H_1[{\bf r}_i^a(z)] \,,
\end{equation}
it is not difficult to show that the variational free energy $F_\rho = -T\ln Z_v +\langle H - H_v\rangle_v$
is given by:
\begin{widetext}
\begin{eqnarray}
F_\rho & = & \sum_{a=1}^p\int dz \int [d{\bf r}(z)]\;\Big\{\,T\rho^a[{\bf r}(z)]\ln\Big(\frac{\rho^a[{\bf r}(z)]}{N}\Big)
+ \Big[\frac{1}{2}\,K\Big(\frac{d{\bf r}}{dz}\Big)^2 + V_{ext}({\bf r}(z))\Big]\rho^a[{\bf r}(z)] \Big\}
\nonumber\\
& + & \frac{1}{2}\sum_{a=1}^p\int dz\int[d{\bf r}(z)][d{\bf r}'(z)]\; \rho^a[{\bf r}(z)]V\big({\bf r}(z) - {\bf r}'(z)\big)\rho^a[{\bf r}'(z)]
\nonumber\\
& - & \frac{1}{2T}\sum_{a,b}\int dz\,dz' \int[d{\bf r}(z)][d{\bf r}'(z')]\; 
\rho^a[{\bf r}(z)]\Delta\big({\bf r}(z) - {\bf r}'(z'),z-z'\big)\rho^b[{\bf r}'(z')]\,.
\end{eqnarray}
Extremizing $F_\rho$ with respect to $\rho^a[{\bf r}(z)]=\rho^a\big({\bf r}(0),\ldots,{\bf r}(Md)\big)$ 
leads to an expression similar to the right hand side of Eq. (\ref{Eq:rho-trajectory}) 
for $\rho^a[{\bf r}(z)]$, with the following expression
for the effective Hamiltonian $H_1$:
\begin{eqnarray}
H_1[{\bf r}(z)] & = & \int dz \Big\{\Big[\frac{1}{2}\,K\Big(\frac{d{\bf r}}{dz}\Big)^2 
+ V_{ext}({\bf r}(z))\Big] + \int[d{\bf r}'(z)]\,V\big({\bf r}(z) - {\bf r}'(z)\big)\rho^a[{\bf r}'(z)]
\nonumber\\
& - & \frac{1}{T} \sum_b\int dz'\int[d{\bf r}'(z')]\; \Delta\big({\bf r}(z)-{\bf r}'(z'),z-z'\big) \rho^b[{\bf r}'(z)]
\Big\}\,.
\label{Eq:resH1-disordered}
\end{eqnarray}
We now write ${\bf r}(z)={\bf r}_0 + {\bf u}(z)$ and Taylor expand the above Hamiltonian 
up to quadratic order in the small displacement field ${\bf u}({\bf r},z)$. Here again we find that,
to order ${\cal O}(u^2)$, $H_1$ can be written in the decoupled form (\ref{H1decomp}),
with (here $\bar\Delta({\bf r})=\int_{-\infty}^\infty dz\,\Delta({\bf r},z)$):
\begin{subequations}
\begin{eqnarray}
H_{1,c.m.}({\bf r}_0) &=& \int dz \int [d{\bf r}'(z)] V\big({\bf r}_0-{\bf r}'(z)\big)\,\rho^a[{\bf r}'(z)] 
-\frac{1}{T}\sum_b\int dz'\int[d{\bf r}'(z')]\,\bar\Delta({\bf r}_0-{\bf r}'(z'),z-z')\rho^b[{\bf r}'(z')] \,,
\label{H1cm-dis}
\nonumber\\
\\
H_{1,u}[{\bf u}(z)] & = & \int dz \,\Big[\frac{1}{2} K\Big(\frac{d{\bf u}}{dz}\Big)^2 + \frac{1}{2}\mu_{\alpha\beta} u_\alpha(z)u_\beta(z)\Big]\,.
\nonumber\\
\label{H1u-dis}
\end{eqnarray}
\end{subequations}
In the above expression of $H_{1,u}[{\bf u}(z)]$, the confining ``mass" tensor is given by:
\begin{eqnarray}
\mu_{\alpha\beta} & = & \int [d{\bf r}'(z)] \,\partial_\alpha\partial_\beta V\big({\bf r}_0-{\bf r}'(z)\big)\rho^a[{\bf r}'(z)] 
-\frac{1}{T}\sum_b\int dz'\int[d{\bf r}'(z')]\,\partial_\alpha\partial_\beta\Delta({\bf r}_0-{\bf r}'(z'),z-z')\rho^b[{\bf r}'(z')] \,,
\label{mualphabeta-dis}
\nonumber\\
\end{eqnarray}
\end{widetext}
and vanishes in hydrodynamics for the same reason as in the undisordered case. Also, like in the pure case, 
the density matrix decouples, $\rho[{\bf r}(z)]=\rho_{c.m.}[{\bf r}_0]\,\rho_u[{\bf u}(z)]$, and hence
the internal modes $\{{\bf u}'\}$ in the expression of $H_{1,c.m.}({\bf r}_0)$ can be integrated out, with the result:
\begin{equation}
H_{1,c.m.}({\bf r}_0) = V_{e}^a({\bf r}_0) + \sum_b \int d{\bf r}_0'
\;\Gamma_{ab}({\bf r}_0-{\bf r}_0')\rho_{c.m.}^b({\bf r}') \,,
\end{equation}
where we introduced an external ``source" potential $V_e^a$ that depends only on the c.m. position ${\bf r}_0$,
and where the kernel $\Gamma_{ab}$ is given by:
\begin{eqnarray}
\Gamma_{ab}({\bf r}_0)=L\Big(\tilde{V}({\bf r}_0)\delta_{ab} 
-\frac{\tilde\Delta({\bf r}_0) }{T} \Big)\,.
\label{Eq:defGamma_ab}
\end{eqnarray}
In the above equation, the potential $\tilde{V}$ is given by Eq. (\ref{Eq:tildeV2}), while $\tilde\Delta$ is
similarly given by:
\begin{eqnarray}
\tilde\Delta({\bf r}) & = & \int[d{\bf u}'(z)]\; \bar\Delta\big({\bf r}-{\bf u}'(z)\big)\,\rho_u^a[{\bf u}'(z)]\,,
\label{Eq:tildeDelta1}
\\
& = & \int_{\bf k} e^{i{\bf k}\cdot{\bf r} - \frac{1}{2}k^2\langle u^2\rangle}\bar\Delta({\bf k})\,.
\label{Eq:tildeDelta2}
\end{eqnarray}
By analogy with Eqs. (\ref{Eq:def_chi_rigid})-(\ref{chi1}), the density response function 
$\chi^{ab}({\bf r},{\bf r}')=-\delta\rho_{c.m.}^a({\bf r}_0)/\delta V_e^b({\bf r}_0')$
satisfies the following equation:
\begin{eqnarray}
\chi^{ab}({\bf r}_0,{\bf r}_0') &=& \beta\rho^a({\bf r})\Big[
\delta_{ab}\delta({\bf r}-{\bf r}') 
\nonumber\\
&-& \sum_c\int d{\bf r}_0''\;\Gamma_{ac}({\bf r}_0-{\bf r}_0'')\chi^{cb}({\bf r}_0'',{\bf r}_0')\Big] \,.
\nonumber\\
\end{eqnarray}
For a homogeneous liquid ($V_e=0$, $\rho({\bf r})=\rho=\mbox{Cst.}$), $\chi^{ab}({\bf r}_0,{\bf r}_0')$ is translationally invariant,
and the above equation can be cast, in Fourier space, into the following matricial form:
\begin{equation}
\sum_c\tilde{\Gamma}^{ac}({\bf q})\chi^{cb}({\bf q}) = \delta_{ab} \,,
\label{Eq:eq_chi_dis}
\end{equation}
with
\begin{equation}
\tilde{\Gamma}_{ac}({\bf q}) = \Big(\frac{T}{\rho}+L\tilde{V}({\bf q})\Big)\delta_{ac} 
- \frac{L\tilde\Delta({\bf q})}{T} \,.
\end{equation}
The matrix $\tilde{\Gamma}$ can easily be inverted using an identity for inverting $p\times p$
matrices of the form:
\begin{equation}
(A^{-1})_{ij} = a\delta_{ij} + b\,,
\end{equation}
namely:
\begin{equation}
A_{ij} = \frac{1}{a}\delta_{ij} - \frac{b}{a(a+pb)}\,.
\end{equation}
For the response function $\chi^{ab}({\bf q})$, this gives in the limit $p\to 0$ the following result:
\begin{eqnarray}
\chi^{ab}({\bf q}) = \frac{1}{L\tilde{V}({\bf q}) + T/\rho}\,\delta_{ab} + 
\frac{L\tilde\Delta({\bf q})}{T\big[L\tilde{V}({\bf q}) + T/\rho\big]^2},
\end{eqnarray}
and hence the diagonal (in replica space) Ursell function 
for the c.m. mode is given by:
\begin{equation}
S_{0,nn}({\bf q}) = \frac{T}{L\tilde{V}({\bf q}) + T/\rho} 
+ \frac{L\tilde\Delta({\bf q})}{\big[L\tilde{V}({\bf q}) + T/\rho\big]^2}\,.
\end{equation}
We thus see that disorder produces a Lorentzian-squared correction to the 
Ursell function fo the c.m. mode in the hydrodynamic limit. The Ursell function of the flux-line
liquid is however unchanged with respect to the pure case (since the confining mass $\mu$ is still zero),
and is given by Eq. (\ref{Eq:res_Snn}), in contrast to the results of refs.  \cite{Nelson-LeDoussal}
and \cite{Tauber-Nelson}.
Note, however, that if a finite mass coefficient $\mu$, then the Ursell function
of the flux liquid in the hydrodynamic limit is given by Eq. (\ref{Eq:S-S0}), 
which in the present context becomes:
\begin{eqnarray}
S_{nn}({\bf q},z) &=&  \Big[\frac{T}{L\tilde{V}({\bf q}) + T/\rho} 
+ \frac{L\tilde\Delta({\bf q})}{\big[L\tilde{V}({\bf q}) + T/\rho\big]^2} - \rho\Big]
\nonumber\\
&\times&\,e^{-q^2\langle u^2\rangle/d_\perp}
+ \rho\exp\Big(-\frac{Tq^2}{2K}|z|\Big) \,.
\label{Eq:S-S0-dis}
\end{eqnarray}

\end{document}